\documentclass[12pt]{article}

\usepackage{epsfig}
\usepackage{amssymb}
\usepackage{amsmath}
\usepackage{color}
\usepackage{enumitem, array}
\usepackage{hyperref}
\usepackage{bbm}
\usepackage{mathrsfs}
\usepackage{colonequals}
\usepackage{cite}
\newcommand{\beqa}{\begin{eqnarray}}
\newcommand{\eeqa}{\end{eqnarray}}
\newcommand{\beq}{\begin{equation}}
\newcommand{\eeq}{\end{equation}}

\makeatletter \@addtoreset{equation}{section} \makeatother

\newdimen\yysquaresize
\newdimen\yyrsquaresize
\newdimen\yythickness
\newdimen\yyskip
\def\yysquare#1{%
\setlength{\yyrsquaresize}{\yysquaresize}%
\addtolength{\yyrsquaresize}{-2\yythickness}%
\vrule width \yythickness%
\vbox to \yysquaresize{%
  \hrule height \yythickness\vss%
  \hbox to \yyrsquaresize{\hss#1\hss}%
  \vss\hrule height \yythickness}%
\vrule width \yythickness}

\def\yyyoung#1{\vtop{\baselineskip0pt\lineskip-\yythickness\halign{\tabskip-\yythickness&\yysquare{##}\cr #1}}}

\newcommand{\young}[1]{\hskip\yyskip\mbox{\yyyoung{#1}}\hskip\yyskip}

\newcounter{multieqs}

\newcommand{\be}{\begin{equation}}
\newcommand{\ee}{\end{equation}}
\newcommand{\bea}{\begin{eqnarray}}
\newcommand{\eea}{\end{eqnarray}}

\def\calN{{\cal N}}

\def\calZ{{\cal Z}}

\def\hws{{\mathfrak{v}}}
\def\g{{\mathfrak{g}}}

\def\Nm{{\mathcal{N}}}
\def\Om{{\mathcal{O}}}

\def\be{\begin{equation}}
\def\ee{\end{equation}}
\def\bea{\begin{eqnarray}}
\def\eea{\end{eqnarray}}

\begin{document}

\thispagestyle{empty}
\setcounter{page}{0}
\begin{flushright}\footnotesize
\texttt{HU-Mathematik-2015-14}\\
\texttt{HU-EP-15/60}\\
\texttt{MITP/15-112}\\
\vspace{0.5cm}
\end{flushright}
\setcounter{footnote}{0}

\begin{center}%
{\Large\textbf{\mathversion{bold}%
On correlation functions of BPS operators in \\$3d$ $\mathcal{N}=6$ superconformal theories}\par}

\vspace{15mm}
{\sc 
 Pedro Liendo$^{a}$,
Carlo Meneghelli$^{b}$,
 Vladimir Mitev$^{a,c}$    }\\[5mm]

{\it $^a$ Institut f\"ur Mathematik und Institut f\"ur Physik,\\ Humboldt-Universit\"at zu Berlin\\
IRIS Haus, Zum Gro{\ss}en Windkanal 6,  12489 Berlin, Germany
}\\[5mm]

{\it $^b$
Simons Center for Geometry and Physics, \\
Stony Brook University, Stony Brook, NY 11794-3636, USA
}\\[5mm]

{\it $^c$ PRISMA Cluster of Excellence,\\
Institut f\"ur Physik, WA THEP,\\
Johannes Gutenberg-Universit\"at Mainz,\\
Staudingerweg 7, 55128 Mainz, Germany
}\\[5mm]

\small{\texttt{pliendo@physik.hu-berlin.de}\\
\texttt{cmeneghelli@scgp.stonybrook.edu}\\
\texttt{vmitev@uni-mainz.de}}\\[20mm]

\textbf{Abstract}\\[2mm]
\end{center}
We introduce a novel harmonic superspace for $3d$ $\calN=6$ superconformal field theories that is tailor made for the study of correlation functions of BPS operators. We calculate a host of two- and three-point functions in full generality and put strong constraints on the form of four-point functions of some selected BPS multiplets. For the four-point function of $\tfrac{1}{2}$-BPS operators we obtain the associated Ward identities by imposing the absence of harmonic singularities.
The latter imply the  existence of a solvable subsector in which the correlator becomes topological.
 This mechanism can be explained by cohomological reduction with respect to a special nilpotent supercharge.

\newpage

\setcounter{tocdepth}{2}
\hrule height 0.75pt
\tableofcontents
\vspace{0.8cm}
\hrule height 0.75pt
\vspace{1cm}

\setcounter{tocdepth}{2}

\section{Introduction and overview}
\label{sec:introduction}

Understanding the general structure of superconformal field theories (SCFTs) has been an important goal in theoretical physics, with many modern developments being motivated by the AdS/CFT correspondence. Although the ``holographic principle'' is expected to be a general feature of conformal field theories, the supersymmetric case offers a unique arena in which many ideas can be tested with a high degree of analytic control. Due to the strong-weak nature of the duality, it would be ideal to have a framework in which aspects of AdS/CFT can be explored without relying on perturbative calculations. Motivated by this, in this work we will study the constraints imposed by superconformal symmetry on correlations functions for the interesting case of $\calN=6$ SCFTs in three dimensions.

In this class of theories falls one of the most remarkable examples of AdS/CFT, namely, the duality between $3d$ $\calN=6$ superconformal Chern-Simons, better known as the ABJ(M) theory, and type IIA string theory on AdS$_4\times \mathbb{CP}^3$ \cite{Aharony:2008ug, Aharony:2008gk}. In many respects ABJ(M) is analogous to the more studied
case of $4d$ $\calN=4$ super Yang-Mills (SYM). It exhibits planar integrability, both at the level of the spectrum of anomalous dimensions, see \cite{Klose:2010ki} for a  review, and
at the level of scattering amplitudes, 
see \cite{Bargheer:2010hn,Huang:2010qy} and references thereafter.
Hence, three-dimensional ABJ(M) parallels four-dimensional $\calN=4$ SYM. It is therefore of great importance to understand 
the constraints that arise from the $\Nm=6$ superconformal algebra. For $\Nm=4$ SYM, this analysis was performed in a series of papers  \cite{Eden:2000bk,Eden:2001ec,Dolan:2001tt,Heslop:2002hp,Dolan:2004mu,Nirschl:2004pa,Dolan:2006ec}, where the Ward identities for correlators of $\frac{1}{2}$-BPS operators were obtained and solved. To the best of our knowledge, there has not been a systematic analysis of the implications of the $\calN=6$ superconformal algebra on correlation functions, and in this paper we attempt to fill this gap in the literature.

A second motivation for our work is the conformal bootstrap idea, in which basic consistency requirements such as unitary and crossing are used to put constraints on the dynamical information of a CFT. In recent years there has been a revival of the bootstrap approach thanks to the numerical techniques developed in \cite{Rattazzi:2008pe}. This approach is non-perturbative in nature and well suited for supersymmetric theories, where the extended symmetry algebra puts strong restrictions on the form of the correlation functions. The $3d$ $\calN=8$ superconformal bootstrap was initiated in \cite{Chester:2014fya}, and in order to extend the bootstrap approach to the $3d$ $\calN=6$ case it is necessary to have a detailed understanding of correlation functions with this symmetry. The techniques developed in this paper help fulfill this goal, and constitute the first step toward the $\Nm=6$ superconformal bootstrap. In a parallel development, it was observed in \cite{Chester:2014mea} (following the work of \cite{Beem:2013sza}), that the $3d$ bootstrap equations with $\Nm \geq 4$ have a solvable truncation in terms of a $1d$ topological theory. In this paper we will rederive this result as a solution of our Ward identities for the case of $\frac{1}{2}$-BPS operators in $\Nm=6$ theories.

Theories with extended supersymmetry can be naturally described using superfields that, in addition to the usual space-time coordinates, depend on Gra\ss mann as well as R-symmetry, or harmonic, coordinates. The spaces of such superfields are known as harmonic superspaces and have been introduced in \cite{Galperin:1984av,Galperin:1985ec}, see also \cite{Galperin:2001uw}.
 Our main purpose in this article is to introduce a new harmonic superspace designed specifically for $3d$ $\calN=6$ theories  and to use it to derive the Ward identites for correlation functions of superconformal multiplets, in particular, short multiplets of the BPS type. 
It should be considered as the analogue of $\mathcal{N}=4$  harmonic/analytic superspace in four dimensions,
 see e.g.~\cite{Doobary:2015gia} and references therein.

Since there have been many works that have studied superconformal symmetry in three dimensions for various values of $\calN$, a brief overview of some relevant results is in order. 
For the maximal number of supersymmetry $\calN=8$, the two-, three-, and four-point functions of $\frac{1}{2}$-BPS operators were described in \cite{Dolan:2004mu} using harmonic superspace. These results are of particular importance because in $\Nm=8$ theories the stress tensor sits in a $\frac{1}{2}$-BPS multiplet, and were used in \cite{Chester:2014fya} to implement the $\Nm=8$ stress-tensor bootstrap.
For the $\Nm=0$ case conserved currents of any spin were studied in \cite{Giombi:2011rz}. Two- and three-point functions of conserved current multiplets, which include the stress tensor as well as flavor currents, were described in \cite{Buchbinder:2015qsa} for $\calN=1,2,3$ and in \cite{Buchbinder:2015wia} for $\calN=4$ using standard Minkowski superspace $\mathbb{M}^{3|2\calN}$. $\calN=6$ multiplets have not been systematically studied yet. Since the proliferation of indices makes Minkowski superspace unsuitable for this task, we propose a superspace approach tailored for BPS multiplets in $\calN=6$ theories. Examples of operators who sit in this type of multiplet include the stress tensor, determinant operators, and monopole operators.

This article is structured as follows. In section \ref{sec:the superconformal algebra} we give a brief overview of the $\Nm=6$ superconformal algebra and its unitary representations. In section \ref{sec: a novel superspace} we present a new superspace  based on a specific decomposition of the superalgebra, optimized for the description of BPS operators. In section \ref{Sec:twoandthreepoints}, we use our superspace to describe the two-point covariant objects that we will use throughout the rest of the article to build correlation functions. We also show how to build three-point functions for several short multiplets. In section \ref{Sec:fourpoints} we study the four-point function of $\frac{1}{2}$-BPS operators and write its associated Ward identity. In particular, this result implies that that in a certain limit the correlator is purely topological. In this section we also present some partial results for the stress-tensor multiplet, which in $\Nm=6$ theories is $\frac{1}{3}$-BPS. In section \ref{sec:cohomological construction} we use a cohomological construction to explain the topological correlator obtained in section \ref{Sec:fourpoints}. We then conclude in section \ref{sec: conclusions}. Most of the technical aspects of our computations are corralled in the appendices.

\section{The superconformal algebra $\mathfrak{osp}(6|4)$
}
\label{sec:the superconformal algebra}

The superconformal algebra $\g=\mathfrak{osp}(6|4)$ is an extension of the
conformal algebra in three dimensions 
 $\mathfrak{sp}(4,\mathbb{R})\simeq \mathfrak{so}(3,2)$.
As any Lie superalgebra, 
 $\g=\mathfrak{osp}(6|4)$ decomposes   into its
bosonic  and  fermionic  parts, $\g=\g_{\bar{0}}\,+\,\g_{\bar{1}}$,
where 
$\g_{\bar{0}}$ is a subalgebra of $\g$ and $\g_{\bar{1}}$ transforms in some representation of $\g_{\bar{0}}$.
In the case of  the superconformal algebra $\mathfrak{osp}(6|4)$ we have 
\be\label{gfermionic_bosonic}
\g_{\bar{0}}=\mathfrak{so}(6)\,\oplus\, \mathfrak{sp}(4)\,, \qquad \g_{\bar{1}}=\underline{6}\,\otimes \underline{4}\,,
\ee
with $\dim\g_{\bar{0}}=15+10=25$ and $\dim\g_{\bar{1}}=6\times 4 =24$. We denote the bosonic generators by $\{\mathfrak{P}_{\alpha\beta}, \mathfrak{K}^{\alpha\beta}, \mathfrak{D}, \mathfrak{R}_{AB}\}$ and the fermionic supercharges by 
$\{\mathfrak{Q}^A_{\alpha}, \mathfrak{S}^A_{\alpha}\}$, where the indices run over $\alpha,\beta=1,2$, $A,B=1,\dots,6$.

Let us quickly review the classification of unitary highest weight representations of $\mathfrak{osp}(6|4)$.
The conformal algebra contains a distinguished generator $\mathfrak{D}$ called dilatation (or energy) operator.
The Lie superalgebra $\mathfrak{osp}(6|4)$ decomposes in eigenspaces with respect to the adjoint action of 
$\mathfrak{D}$:
\bigskip
\newline
\begin{tabular}{%
l
                >{\centering }m{2cm}
     >{\centering }m{2cm}
     >{\centering }m{2.3cm}
 >{\centering }m{2cm}
             >{\centering\arraybackslash}m{2cm}}
\hline             
& 
$\mathfrak{P}_{\alpha\beta}$
&
$\mathfrak{Q}^A_{\alpha}$
&
$\mathfrak{D}$, $\mathfrak{L}^{\alpha}_\beta$, $\mathfrak{R}_{AB}$
&
$\mathfrak{S}^A_{\alpha}$
&
$\mathfrak{K}^{\alpha\beta}$
\\\hline
Eigenvalue of $[\mathfrak{D},\cdot]$
&
$+1$
&
$+\frac{1}{2}$
& 0
&$-\frac{1}{2}$
& $-1$
\\
\end{tabular}
\bigskip
\newline
The states with the lowest value of dilatation weight within a unitary irreducible representation 
of $\mathfrak{osp}(6|4)$ are called primary states. From the eigenvalues of $\mathfrak{D}$, it follows that primary states are annihilated by the the generators with negative weight: $\mathfrak{S}^{\alpha}_A$ and $\mathfrak{K}^{\alpha\beta}$.
Moreover, they transform in some finite dimensional representation of the zero weight part 
$\mathfrak{u}(1)\oplus \mathfrak{su}(2)\oplus\mathfrak{so}(6)$ generated by 
$\mathfrak{D}$, $\mathfrak{L}^{\alpha}_\beta$, and $\mathfrak{R}_{AB}$.
We label such representations by  the dilatation weight $\Delta$, the spin $s\in 
\frac{1}{2}\,\mathbb{Z}_{\geq 0}$ and  the $\mathfrak{so}(6)$ Dynkin labels $(r_1,r_2,r_3)$.
These data uniquely specify a unitary representation of $\mathfrak{osp}(6|4)$ and we group them as follows:
\beq\label{replabale_1}
\{(r_1,r_2,r_3),(\Delta,s)\}\,.
\eeq
Apart from the one dimensional representation with $\{(r_1,r_2,r_3),(\Delta,s)\}=\{(0,0,0),(0,0)\}$,
 unitary irreducible representations of $\mathfrak{osp}(6|4)$ 
satisfy the following unitarity bounds:
\begin{align}
\Delta=r_1&\qquad \text{when}\,\,\,\,\,\,s=0\,,\\
\Delta\geq s+r_1+1&\qquad \text{when}\,\,\,\,\,\,s\geq 0\,.
\end{align}
The first case corresponds to isolated short representations,
see
\cite{Minwalla:1997ka,Bhattacharya:2008zy}.
These are distinguished by the fact that they cannot recombine with other short representations.

It can be useful to compare this result with the non-supersymmetric case.
Unitary irreducible representations of $\mathfrak{sp}(4,\mathbb{R})$, 
other than the trivial representation, 
satisfy the following unitarity bounds:
\begin{align}
\Delta\geq\tfrac{1}{2}&\qquad \text{when}\,\,\,\,\,\,s=0\\
\Delta\geq 1&\qquad \text{when}\,\,\,\,\,\,s=\tfrac{1}{2}\\
\Delta\geq s+1 &\qquad \text{when}\,\,\,\,\,\,s\geq 1
\end{align}
The bounds are saturated by massless boson, massless fermion, and conserved higher-spin currents respectively, 
see \cite{Dirac:1963ta,Flato:1978qz}.
For $\mathfrak{osp}(\mathcal{N}|4)$ with $\mathcal{N}\geq 1$, the massless boson and fermion combine 
into a single representation.

\begin{table}
\centering
\renewcommand{\arraystretch}{1.6}
\begin{tabular}{%
| l
                |>{\centering }m{3.1cm}
 |>{\centering }m{3.1cm}
             |>{\centering\arraybackslash}m{5.5cm}|
}
\hline
Type 
&Name 
& Fraction of $\mathfrak{Q}$-susy that kills the h.w.s.
&$\{(r_1,r_2,r_3),(\Delta,s)\}$
\\\hline
Short
& 
$(3,B,1)$ \linebreak 
$(3,B,2)$ \linebreak
$(3,B,\pm)$
&
$1/6$ \linebreak 
$1/3$ \linebreak
$1/2$
&
$\{(k,0,0),(k,0)\}$ \linebreak 
$\{(k,k,0),(k,0)\}$ \linebreak
$\{(\tfrac{1}{2}k,\tfrac{1}{2}k,\pm \tfrac{1}{2} k),(\tfrac{1}{2}k,0)\}$
\\\hline
Semi-Short
& 
$(3,A,1)$ \linebreak 
$(3,A,2)$ \linebreak
$(3,A,\pm)$
&
$1/12$ \linebreak 
$1/6$ \linebreak
$1/4$
&
$\{(k,0,0),(k+s+1,s)\}$ \linebreak 
$\{(k,k,0),(k+s+1,s)\}$ \linebreak
$\{(\tfrac{1}{2}k,\tfrac{1}{2}k,\pm \tfrac{1}{2}k),(\tfrac{1}{2}k+s+1,s)\}$
\\\hline
Cons. Currents
& 
$(3,\,\text{cons})$
&
$1/3$
&
$\{(0,0,0),(s+1,s)\}$
\\\hline
\end{tabular}
\renewcommand{\arraystretch}{1.0}
\caption{Non-long representations of $\mathfrak{osp}(6|4)$ in the notation of 
 \cite{Dolan:2008vc}. The first label is fixed as $3=\mathcal{N}/2$.
The indices run as $k\in\mathbb{Z}_{>0}$ and $2s\in\mathbb{Z}_{\geq0}$.}
\label{tab:nonlong reps}
\end{table}
As it will become clear in the following sections, it is convenient to reshuffle the representation labels \eqref{replabale_1} as
\beq\label{new_Osp_labels}
\big{\{}(b_+,b_-),(s,j,c_{2|2})\big{\}}=
\big{\{}\left(\tfrac{\Delta}{2}+\tfrac{r_1+r_2}{4}+r_3,\tfrac{\Delta}{2}+\tfrac{r_1+r_2}{4}-r_3\right),\left(s,r_1-r_2,\Delta-\tfrac{r_1+r_2}{2}\right)\big{\}}\,.
\eeq 
In the equation above, $(s,j,c_{2|2})$ label a finite dimensional irreducible representation of $\mathfrak{su}(2|2)$ where $(j,s)$ are spin labels of  the
 $\mathfrak{su}(2)\oplus \mathfrak{su}(2)$  subalgebra, i.e. the eigenvalues of $\mathfrak{L}^{1}_{1}$ and $\mathfrak{R}^{1}_{1}$ on the highest weight state, and $c_{2|2}$ is the eigenvalue of the central charge generator $\mathfrak{C}$. We refer to appendix \ref{app: Algebra conventions from oscillator realization } for more details.
It is remarkable that finite dimensional representations of $\mathfrak{su}(2|2)$ admit a continuous label $c_{2|2}$.
Multiplets with label $b_-=0$, respectively $b_+=0$, will be called \textit{chiral}, respectively \textit{anti-chiral}.
 Finally, we define the conjugate representation as 
\beq
\label{conjugation}
\big{\{}(b_+,b_-),(s,j,c_{2|2})\big{\}}^*=\big{\{}(b_-,b_+),(s,j,c_{2|2})\big{\}}\,.
\eeq

Of particular importance is the $\mathcal{N}=6$ stress-tensor supermultiplet, denoted by $(3,B,2)$ in Table \ref{tab:nonlong reps}, which contains the stress-tensor operator and the $\mathfrak{so}(6)$ R-symmetry conserved currents.
This multiplet can be found in 
\cite{Nilsson:1984bj}, see also \cite{Papathanasiou:2009en}.
As pointed out in \cite{Bashkirov:2011fr}, it also contains an additional $\mathfrak{u}(1)$ conserved current. Table \ref{tab:TN=6} in appendix \ref{App:covent} contains the decomposition of this supermultiplet in representations of the bosonic subalgebra $\mathfrak{so}(6)\oplus\mathfrak{sp}(4)$.

\section{A novel superspace}
\label{sec: a novel superspace}

In the following we will introduce a superspace\footnote{A different three-dimensional harmonic superspace has already been used to study the ABJM theory, see \cite{Buchbinder:2008vi, Buchbinder:2009dc}.  The purpose of that superspace is to define the theory off-shell and develop a supergraph perturbation theory where the maximal amount of supersymmetry is manifest.  It is thus very different from the approach of our paper. } tailored to describe BPS multiplets. Multiplets of this type include the $(3,B,2)$ family which is $\frac{1}{3}$-BPS and for $k=1$ corresponds to the stress-tensor multiplet; and $(3,B,\pm)$ multiplets which are $\frac{1}{2}$-BPS and can be represented as superfields satisfying some sort of chirality condition. More general supermultiplets are described by superfields with indices. This is a super-analog of the description of representations of the conformal group  with non-trivial spin in  ordinary Minkowski space.

\paragraph{Graded decomposition of $\mathfrak{osp}(6|4)$.}
Let us consider a $\mathbb{Z}\times \mathbb{Z}$ weight decomposition\footnote{
We are using the notation $+$ for the direct sum of super-vector spaces and $\oplus$ for the direct sum of Lie superalgebras.
}
\beq\label{ZZgraded}
\mathfrak{osp}(6|4)\,=\,\g_{(-,-)}\,+\,\g_{(-,0)}\,+\g_{(0,-)}\,+\g_{(0,0)}\,+\g_{(+,0)}\,+\g_{(0,+)}\,+\,\g_{(+,+)}\,,
\eeq
 such that the highest weight state of the representations of type $(3,B,2)$
transforms in a one-dimensional representation of $\g_{(0,0)}$ and is annihilated by the negative part of $\g$.
The graded decomposition \eqref{ZZgraded} corresponds to eigenspaces with respect to the action of 
$\mathfrak{C}_{L/R}=\mathfrak{D}+\mathfrak{J}_{L/R}$, where $\mathfrak{D}$ is the dilatation generators and 
$\mathfrak{J}_{L/R}$ belong to the Cartan subalgebra of the $\mathfrak{so}(6)$ R-symmetry, 
see \eqref{JLRdef} for their explicit expressions.
 As in \eqref{gfermionic_bosonic}, it follows from 
\beq\label{gg_gradingrespected}
[\g_{(a,b)},\g_{(c,d)}\}\,\simeq\,
\g_{(a+c,b+d)}\,,
\eeq
that each component $\g_{(a,b)}$ transforms in some representation of $\g_{(0,0)}$.
In the case of  \eqref{ZZgraded} we have that
\beq 
\g_{(0,0)}\,=\,\mathfrak{gl}(2|2)\,\oplus\,\mathfrak{gl}(1)\,,
\eeq
and
\beq\label{gpositivetransf}
\g_{(+,0)}\simeq \young{\cr}_{(+,0)}\,,
\quad
\g_{(0,+)}\simeq \young{\cr}_{(0,+)}\,,
\quad
\g_{(+,+)}\simeq \young{& \cr }_{(+,+)}\,,
\eeq
where $\young{\cr}$ and  $\young{& \cr }$  respectively 
denote the fundamental  and the graded symmetric representations of $\mathfrak{sl}(2|2)$.
The negative part in  \eqref{ZZgraded}  transforms in the representation  conjugate to \eqref{gpositivetransf}.
More details can be found in Appendix \ref{App:covent}.

\noindent
\emph{Remark:} The transformation properties \eqref{gpositivetransf} of the positive part of $\mathfrak{osp}(6|4)$ resemble
the transformation properties of the so-called magnon excitations of the ABJM spin-chain,
see \cite{Zarembo:2009au} and references therein.
This is of course not a coincidence as the  vacuum of the length $L$ spin-chain 
is the highest weight state of the representation $(3,B,1)$ with $k=L$.
Only the excitations corresponding to $\g_{(+,0)}$ and $\g_{(0,+)}$
should be regarded as fundamental, the one associated to $\g_{(+,+)}$ can 
be obtained as commutators of the former. 

\paragraph{The supergroup $OSP(6|4)$.}
We define the  orthosymplectic group  as 
\beq\label{OSPgroupdef}
OSP(6|4)\,=\,
\Big{\{}
g\,\in\,GL(6|4)\,\,
\text{such that }\,
g^{st}\,\eta\, g\,=\,\eta
\Big{\}}\,.
\eeq
In the equation above $st$ denotes super-transposition (an operation with the properties 
$(AB)^{st}=B^{st} A^{st}$ and $(A^{st})^{st}=\Pi\,A\,\Pi$ where $\Pi$ is the 
super-parity matrix which acts as $+1$ on bosons and $-1$ on fermions)  and $\eta$ is a supersymmetric matrix, 
i.e.~$\eta=\eta^{st}\Pi=\Pi\,\eta^{st}$.
With the definition \eqref{OSPgroupdef}, the center $\mathcal{Z}(OSP(6|4))=\{+\mathbbm{1},-\mathbbm{1}\}\simeq \mathbb{Z}_2$ is a subgroup of $\mathcal{Z}(SO(6)\times SP(4))=\{+\mathbbm{1},-\mathbbm{1},\Pi\}$.
We further denote the special orthosymplectic group, i.e. the subgroup of elements in $OSP(6|4)$
with unit superdeterminant, as $SOSP(6|4)$.

Next, we  choose our conventions for $\eta$ and $\Pi$ such that the  decomposition \eqref{ZZgraded} 
of the algebra corresponds to 
a triangular decomposition of $OSP(6|4)$.
This is achieved by taking the super-parity matrix $\Pi$ 
and the super-symmetric matrix $\eta$
to be 
\beq
\label{eq:defPiandSigma}
\Pi\,=\,
\begin{pmatrix}
\Sigma & 0 & 0 \\
0 & +\mathbbm{1}_2 & 0 \\
0 & 0  &\Sigma
\end{pmatrix}\,,
\qquad
\eta\,=\,
\begin{pmatrix}
0 & 0 & \mathbbm{1}_4 \\
0 & -\sigma_1 & 0 \\
\Sigma& 0  &0
\end{pmatrix}\,,
\eeq
where
 $\Sigma=\left(\begin{smallmatrix}-\mathbbm{1}_2&0\\0&+\mathbbm{1}_2\end{smallmatrix}\right)$
and
 $\sigma_1=\left(\begin{smallmatrix}0&1\\1&0\end{smallmatrix}\right)$.
Finally, 
we define super-transposition as $(A^{st})_{ij}:=(-1)^{(|i|+1)|j|}\,A_{ji}$, 
where $(-1)^{|i|}:=\Pi_{ii}$. Notice that this definition can be applied to square matrices as well as to rectangular ones. 

\paragraph{Coset superspace.}
We introduce the  superspace  $\mathscr{M}_{(8|8)}$ as a coset
\beq
\mathscr{M}_{(8|8)}\,\simeq\, OSP(6|4)\,\big{/}\,G_{\leq 0}\,,
\eeq
where $G_{(a,b)}$ denotes the  exponentiation of the $\g_{(a,b)}$ subspace of $\mathfrak{osp}(6|4)$, see \eqref{ZZgraded}.
We take $E(\mathbf{p})\in G_{>0}$ as coset representative explicitly given by
\beq
\label{eq:definitionofcosetrepresentative}
E(\mathbf{p}):=
\exp\text{\footnotesize{$
\begin{pmatrix}
0_4  & \bar{W} & W & X \\
0 & 0& 0 & W^{st} \\
0 & 0 & 0 & \bar{W}^{st}\\
0 & 0 & 0 &  0_4
\end{pmatrix}$}}\,=\,
\text{\footnotesize{$\begin{pmatrix}
\mathbbm{1}_4  & \bar{W} & W & X_+ \\
0 & 1 & 0 & W^{st} \\
0 & 0 & 1 & \bar{W}^{st}\\
0 & 0 & 0 &  \mathbbm{1}_4  
\end{pmatrix}$}}\,,
\eeq
where $X_+\colonequals X+\frac{\bar{W}\,W^{st}+W\,\bar{W}^{st}}{2}$,
$\mathbf{p}$ is a point in $\mathscr{M}_{(8|8)}$ and $X$ is a \textit{graded-antisymmetric matrix}, i.e.
\beq
\label{eq:defantisym}
X^{st}=-\Sigma\,X\,.
\eeq
More explicitly, the superspace coordinates  are given by the collection\footnote{
The matrix $x^{\alpha\beta}$ can be written in terms of $x^{\mu}$ as $x^{\alpha\beta}=\left(\begin{smallmatrix}x^0-x^1&x^2\\x^2&x^0+x^1\end{smallmatrix}\right)$.
}
\beq
\label{eq:coordexpl}
X=\left(X^{\mathsf{A}\mathsf{B}}\right)\,=\,
\begin{pmatrix}
x^{\alpha\beta} & \theta^{a\beta} \\
\theta^{b\alpha} & y^{ab}
\end{pmatrix}\,,
\qquad
W\,=\,\left(W^{\mathsf{A}}\right)\,=\,
\begin{pmatrix}
\xi^{\alpha} \\ v^{a} 
\end{pmatrix}\,,
\quad
\bar{W}\,=\,\left(\bar{W}^{\mathsf{A}}\right)\,=\,
\begin{pmatrix}
\bar{\xi}^{\alpha} \\ \bar{v}^{a} 
\end{pmatrix}\,,
\eeq
where $\alpha,\beta=1,2$, $a,b=1,2$, and  $x^{\alpha\beta}=x^{\beta\alpha}$, $y^{ab}=-y^{ba}$. When convenient we will write $y^{ab}=\epsilon^{ab}\,y$.
Notice that there  are  eight bosonic coordinates ($x,y,v,\bar{v}$) and eight fermionic coordinates ($\theta,\xi,\bar{\xi}$). 

The action of $g\,\in\,OSP(6|4)$ on $\mathbf{p}\in \mathscr{M}_{(8|8)}$  is defined by the relation
\beq\label{coset_action}
g\,E(\mathbf{p})\,=\,E(g\circ \mathbf{p})\,g^{*}_{\mathbf{p}}\,,
\qquad
g^{*}_{\mathbf{p}}\,\in\,G_{\leq 0}\,.
\eeq
Notice that the elements  $\pm g$ act in the same way in the superspace $\mathscr{M}_{(8|8)}$,  
so the coordinates $(X,W,\bar{W})$ transform non-trivially  only under the projective supergroup $POSP(6|4)$.
The compact relation \eqref{coset_action}
 encodes the transformation properties $\mathbf{p}\mapsto g\circ \mathbf{p}$ in a rather implicit form.
In order to present the group action more explicitly, it is convenient to introduce
\beq
\mathcal{X}_{\pm}\,:=\,X\,\pm\,\tfrac{1}{2}\,\bar{W}\wedge W\,,
\qquad
\bar{W}\wedge W\,:=\bar{W}W^{st}-W\bar{W}^{st}\,.
\eeq  
\begin{table}
\centering
\renewcommand{\arraystretch}{1.6}
\begin{tabular}{%
| l
                |>{\centering }m{2.30cm}
     |>{\centering }m{2.30cm}
        |>{\centering }m{2.8cm}
 |>{\centering }m{2.35cm}
             |>{\centering\arraybackslash}m{2.35cm}|
}
\hline
& 
$G_{(+,0)}$
&
$G_{(0,+)}$
&
$\widetilde{G}_{(0,0)}$
&
$G_{(-,0)}$
&
$G_{(0,-)}$
\\\hline
\footnotesize{$\!g\circ (\mathcal{X}_+,W)\!\!$}
&
\footnotesize{$ \!(\mathcal{X}_+,W+b)\!$}
&
\footnotesize{$\!(\mathcal{X}_++\bar{b}\,\wedge\, W,W)\!$}
&
\footnotesize{$(A\mathcal{X}_+A^{st},A Wa^{+1})\!\!$}
&
\footnotesize{$(h\,\mathcal{X}_+h^{st},h\, W)$}
&
\footnotesize{$(\mathcal{X}_+,W\!-\!\mathcal{X}_+\bar{c}^{st})\!\!\!$}
\\\hline
\footnotesize{$\!g\circ (\mathcal{X}_-,\bar{W})\!\!$}
&
\footnotesize{$\!(\mathcal{X}_-\!-\bar{W}\,\wedge\, b,\bar{W})\!$}
&
\footnotesize{$\!(\mathcal{X}_-,\bar{W}+\bar{b})\!$}
&
\footnotesize{$(A\mathcal{X}_-A^{st},A\bar{W}a^{-1})\!\!$}
&
\footnotesize{$(\mathcal{X}_-,\bar{W}\!-\!\mathcal{X}_-c^{st})\!\!\!$}
&
\footnotesize{$(\bar{h}\,\mathcal{X}_-\bar{h}^{st},\bar{h}\, \bar{W})$}
\\\hline
\end{tabular}
\renewcommand{\arraystretch}{1.0}
\caption{
Transformation properties under the elements listed in \eqref{Groups_triang:APP}.
We used the definitions 
$h:=(\mathbbm{1}_4+W c)^{-1}$ and
$\bar{h}:=(\mathbbm{1}_4+\bar{W}\bar{c})^{-1}$.
The transformations corresponding to $G_{(+,+)}$, $G_{(-,-)}$  are generated by the ones above.
For completeness we give their explicit form in \eqref{CtransfApp1}, \eqref{CtransfApp2}.}
\label{tab:Xpmtransf}
\end{table}
The pairs $p=(\mathcal{X}_+,W)$, $\bar{p}=(\mathcal{X}_-,\bar{W})$ transform independently under 
the special orthosymplectic supergroup $SOSP(6|4)$, see Table \ref{tab:Xpmtransf}. Superfields that
depend only on $(\mathcal{X}_+,W)$, respectively $(\mathcal{X}_-,\bar{W})$, will be called \textit{chiral}, respectively \textit{anti-chiral}. These naming conventions are consistent with the ones above \eqref{conjugation}. 

The $SOSP(6|4)$ supergroup can be enlarged to 
$OSP(6|4)$ by including the element
\beq
P:=\begin{pmatrix}
\mathbbm{1}_4 & 0 & 0 \\
0 & \sigma_1 & 0 \\
0 & 0  & \mathbbm{1}_4
\end{pmatrix}\,,
\eeq 
with $\text{sdet}(P)=-1$ and $P^2=\mathbbm{1}$.
 Its action on the superspace $\mathscr{M}_{(8|8)}$ is given by
 \beq
 P\circ(\mathcal{X}_+,W)\,=\,(\mathcal{X}_-,\bar{W})\,.
 \eeq
Finally, we may notice that $\eta$ is an element of $OSP(6|4)$ with $\text{sdet}(\eta)=-1$.
 Its action on the superspace is given by
\beq\label{acion_eta}
 \eta\circ(\mathcal{X}_+,W)=(\mathcal{X}_-^{-1}\Sigma,\mathcal{X}_-^{-1}\bar{W})\,,
 \qquad
  \eta\circ(\mathcal{X}_-,\bar{W})=(\mathcal{X}_+^{-1}\Sigma,\mathcal{X}_+^{-1}W)\,.
\eeq
The square of this operation is the transformation generated by $\Pi$, 
which  acts on  $\mathscr{M}_{(8|8)}$
as $(-1)^F$.
The action of  $\eta$ looks like a super-extension of conformal inversion.
We define superconformal inversion as $I:=P\circ\eta$;
this  is  an element of  $SOSP(6|4)$.

\paragraph{Superfields in superspace.}

As explained in \cite{Heslop:2003xu} in a similar context, 
long representations of the superconformal group  can be described 
as so-called quasi-tensor superfields in the  superspace  $\mathscr{M}_{(8|8)}$.
These are fields
\beq\label{long_multi}
\mathcal{O}_{\underline{\mathsf{A}}}(\mathbf{p})\,,
\eeq
where $  \mathbf{p}\in \mathscr{M}_{(8|8)}$ and $\underline{\mathsf{A}}$ runs over  a finite dimensional 
irreducible representation $\rho$ of $\mathfrak{su}(2|2)\,\subset\,\g_{(0,0)}$ with labels  $(s,j,c_{2|2})$, see \eqref{new_Osp_labels}.
It can be verified using \eqref{new_Osp_labels} and the content of Table \ref{tab:nonlong reps}
that for $\frac{1}{2}$-BPS and $\frac{1}{3}$-BPS multiplets such a representation is trivial, so that the corresponding superfield
has no indices $\underline{\mathsf{A}}$. 

The transformation properties of the primary superfield \eqref{long_multi} can be represented as follows.
The group action on the coset
\eqref{coset_action} provides a map
\begin{align}\label{fromSOSPtoSU22}
\Big{(}SOSP(6|4),\,\mathscr{M}_{(8|8)}\Big{)}\,&\rightarrow\,\,
\Big{(}SU(2|2),\,U(1),\,U(1)\Big{)}\,, \\
\Big{(}g,\,\mathbf{p}\Big{)}\,&\mapsto\,\,
\Big{(}M_{\mathbf{p}}(g),\,\omega_{p}(g),\,\bar{\omega}_{\bar{p}}(g)\Big{)}\,.
\end{align}
This map essentially corresponds to taking the block diagonal entries of the matrix $g^*_{\mathbf{p}}$ in the 
definition \eqref{coset_action} of the group action on the coset elements. Specifically, using the parametrization \eqref{Hdef:app} of $g^*_{\mathbf{p}}$, we can write the matrix $H_\mathbf{p}(g)=M_\mathbf{p}(g)\text{diag}(\lambda^{-1},\lambda^{-1}|\lambda, \lambda)$ with $\lambda=\text{sdet}({H}_{\mathbf{p}}(g))^{\frac{1}{4}}$, so that $\text{sdet}(M_\mathbf{p}(g))=1$. Furthermore, we have
\beq
\omega_p(g)\colonequals \sqrt{\text{sdet}({H}_{\mathbf{p}}(g))k_{\mathbf{p}}(g)^{-1}}\,,\qquad \bar{\omega}_{\bar{p}}(g)\colonequals\sqrt{\text{sdet}({H}_{\mathbf{p}}(g))k_{\mathbf{p}}(g)}\,.
\eeq
The most important cases involve the transformations $g\in G_{(-,0)}$ and $g\in G_{(0,-)}$, in which we have from \eqref{eq:gstarforG-0} and \eqref{eq:gstarforG0-}
\beq
\label{eq:actionofomega}
\begin{split}
&g\in G_{(-,0)}\quad \Longrightarrow\quad  \omega_p(g)=\text{sdet}(h_p)\,,\qquad \bar{\omega}_{\bar{p}}(g)=1\,,\\
&g\in G_{(0,-)}\quad \Longrightarrow\quad  \omega_p(g)=1\,,\qquad \bar{\omega}_{\bar{p}}(g)=\text{sdet}(\bar{h}_{\bar{p}})\,,
\end{split}
\eeq
where $h$ and $\bar{h}$ were defined in Table~\ref{tab:Xpmtransf}.
The elements $M,\omega,\bar{\omega}$ 
enter the transformation properties of the
super-primary operator \eqref{long_multi} as follows:
 \beq\label{Otransf}
\mathcal{O}_{\underline{\mathsf{A}}}(\mathbf{p})\,\mapsto\,
\omega_p(g)^{b_+}\,\bar{\omega}_{\bar{p}}(g)^{b_-}\,
\left[\rho(M_{\mathbf{p}}(g))\right]_{\underline{\mathsf{A}}}^{\underline{\mathsf{B}}}\,
\mathcal{O}_{\underline{\mathsf{B}}}(g\circ \mathbf{p})\,.
\eeq

\noindent
\emph{Remark:} Concerning the description of $\frac{1}{2}$-BPS representation in superspace, 
it might seem more natural to consider a different  coset where the $G_0$ part includes the factor $SU(3|2)$ 
that leaves invariant the highest weight state.
This construction would lead to a $6|6$ dimensional superspace also used in 
\cite{Huang:2010qy}  in order
to discuss the so-called dual-superconformal symmetry of ABJM scattering amplitudes.
This choice does not appear to be  particularly efficient in the discussion of correlation functions as 
the
$\frac{1}{2}$-BPS super-multiplets $(3,B,+)$ and $(3,B,-)$ correspond to two different 
$SU(3|2)\subset OSP(6|4)$. 
This subgroup emerges naturally in our construction with
$\g_{(+,0)}+\g_{(0,0)}+\g_{(-,0)}\simeq \mathfrak{gl}(3|2)$ 
and
$\g_{(0,+)}+\g_{(0,0)}+\g_{(0,-)}\simeq \mathfrak{gl}(3|2)$ .

\paragraph{R-symmetry coordinates from  an embedding space}
The internal R-symmetry coordinates $(y,v^a,\bar{v}^a)$, see 
\eqref{eq:coordexpl}, may look exotic at first sight.
We will now elucidate their origin from the perspective of an embedding space.
According to the definition in Section \ref{sec:the superconformal algebra}, primary states
transform in some irreducible representation of the R-symmetry $\mathfrak{so}(6)$.
In the case of the multiplets $(3,B,\pm)$ and $(3,B,2)$ these representations correspond   respectively
to totally symmetric 
tensors in the fundamental/anti-fundamental   $\mathfrak{su}(4)$ indices and mixed tensors subject to a tracelessness condition.
To describe such representations in an index free notation,
one introduces auxiliary variables $V^I,\bar{V}_I$ with $I=1,\dots,4$ subject to the relation $\sum_{I}V^I\bar{V}_I=0$ in order to implement 
the tracelessness condition.
More explicitly, we can define by slight abuse of notation the following expression for the $\tfrac{1}{2}$- and $\tfrac{1}{3}$-BPS operators
\beq
\label{eq:Rharmonicfields}
\begin{split}
\varphi_k(x,V)\colonequals V^{I_1}\cdots V^{I_k}\,\varphi_{I_1\cdots I_k}(x)&\,,\qquad
\bar{\varphi}_k(x,\bar{V})\colonequals\bar{V}_{J_1}\cdots \bar{V}_{J_k}\,\bar{\varphi}^{J_1\cdots J_k}(x)\,,\\
\mathcal{W}_k(x,V,\bar{V})\colonequals V^{I_1}&\cdots V^{I_k}\,\bar{V}_{J_1}\cdots \bar{V}_{J_k}\,\mathcal{W}^{J_1\cdots J_k}_{I_1\cdots I_k}(x)\,.
\end{split}
\eeq
All such fields have homogeneous dependence on  the variables  $V$ and $\bar{V}$.
This scale redundancy, together with the orthogonality condition $\sum_{I}V^I\bar{V}_I=0$, can be respectively gauge fixed and solved by writing
\begin{equation}
\label{eq:oldrelationvwithyandomega}
 V^I\,=\,(1,v^a,y_+)\,,
\qquad
 \bar{V}_I\,=\,(-y_-,-\epsilon_{ab}\bar{v}^b,1)\,,
\end{equation}
with $y_{\pm}\colonequals y\pm\tfrac{1}{2}v^a \epsilon_{ab}\bar{v}^b$.  We remark that setting $y=v=\bar{v}=0$ in \eqref{eq:Rharmonicfields} leaves only the R-symmetry highest weight state.

Hence, we see how one can recover the R-symmetry coordinates of the  harmonic superspace introduced in this section.
It can be of use to recall the 
 main examples  of such representations in terms of composite gauge-invariant operators.
In ABJM, we have the following explicit expression for the $\tfrac{1}{3}$-BPS operators
\begin{equation}
 \mathcal{W}_k(x,V,\bar{V})=
\text{tr}\left[\left(
V\cdot \phi(x)\,\bar{\phi}(x)\cdot\bar{V}\right)^k\right]\,,
\end{equation}
where $V\cdot \phi(x):=V^{I}\phi_{I}(x)$, 
$\bar{\phi}(x)\cdot\bar{V}:=\bar{\phi}^{J}(x)\bar{V}_J$
and the trace is over the gauge group indices.
Correlation functions of such operators have been computed in perturbation theory in  \cite{Young:2014lka,Young:2014sia}
and in 
\cite{Hirano:2012vz} using holography.  
The $\frac{1}{2}$-BPS operators have two type of gauge theory manifestation: as determinant operators and as monopole operators.   
The determinant operators take the schematic form 
\beq
\varphi_k(x)\,=\,
\epsilon_{g_1\dots g_N}\,
\epsilon^{\bar{g}_1\dots \bar{g}_N}\,
V\cdot \phi^{g_1}_{\bar{g}_1}(x)\,\cdots\,
V\cdot \phi^{g_N}_{\bar{g}_N}(x)\,,
\eeq
where $g_i$, $\bar{g}_i$ are indices of the bifundamental representation of the gauge group.
Such operators have been studied in  \cite{Murugan:2011zd,Berenstein:2008dc,Giovannoni:2011pn}. 
Monopole operators on the other hand are not defined as composite operators  but, after using the operator/state correspondence,
by quantizing  certain classical solutions, see  \cite{Berenstein:2009sa}.
They play a crucial role in the enhancement from $\Nm=6$ to $\Nm=8$ superconformal symmetry 
\cite{Bashkirov:2010kz,Benna:2009xd,Gustavsson:2009pm,Kwon:2009ar},
and contribute to the  superconformal index \cite{Kim:2009wb}.

\paragraph{Reduction to $\mathcal{N}=4$ superspace.}
We conclude this section by describing how $\mathcal{N}=6$ superspace and superfields can be decomposed 
with respect to  the $\mathcal{N}=4$ superconformal algebra $\mathfrak{osp}(4|4)$.
All the embeddings of $\mathfrak{osp}(4|4)$ into $\mathfrak{osp}(6|4)$ are equivalent.
A convenient embedding is described in the following.
The relation \eqref{gg_gradingrespected} implies that
\beq\label{convemedding}
\g_{\text{red}}:=
\g_{(-,-)}\,+\,\g_{(0,0)}\,+\,\g_{(+,+)}\,
\eeq
forms a subalgebra of $\mathfrak{osp}(6|4)$.
With a little inspection one can verify that $\g_{\text{red}}\simeq \mathfrak{osp}(4|4)+\mathfrak{u}(1)$,
where the $\mathfrak{u}(1)$ factor corresponds to the parameter $a$ in Table \ref{tab:Xpmtransf}.
Eigenspaces of this $\mathfrak{u}(1)$ transform irreducibly under $\g_{\text{red}}$.
As an example let us consider an $\mathcal{N}=6$ chiral superfield, we have a  decomposition 
\beq\label{N=6toN=4}
\varphi_{k}(\mathcal{X}_+,W)\,=\,
\sum_{n=0}^{k}\,
\varphi^{(n)}_{k,\,\mathsf{A}_1\dots\mathsf{A}_n}(\mathcal{X}_+)\,W^{\mathsf{A}_1}\cdots W^{\mathsf{A}_n}\,,
\eeq
for which each factor $\varphi^{(n)}_{k,\,\mathsf{A}_1\dots\mathsf{A}_n}(\mathcal{X}_+)$ transforms into itself under   
$\mathcal{N}=4$ superconformal transformations.
The decomposition of non-chiral superfields is more subtle as one needs to take into account the $\mathcal{N}=4$ 
super-descendants entering \eqref{N=6toN=4}. 
This is a familiar situation already appearing when decomposing  non-chiral operators from  $\mathcal{N}=1$ to $\mathcal{N}=0$ in four dimensions, see e.g.~\cite{Li:2014gpa}.
The subspace of harmonic superspace obtained by setting $W=\bar{W}=0$ carries the action of the $\mathcal{N}=4$
 superconformal group $OSP(4|4)$ and so does the corresponding superfield.
 This is true for any superfield but it captures only one $\mathcal{N}=4$ supermultiplet
 in the $\mathcal{N}=6\rightarrow \mathcal{N}=4$ decomposition.

\section{Two- and three-point functions of BPS operators}
\label{Sec:twoandthreepoints}

In this section, we will describe the structures appearing in the two- and three-point functions of general primary operators as well as the specific forms of the correlation functions of $\frac{1}{2}$- and $\frac{1}{3}$-BPS operators. As in usual conformal field theory, we find that the two-point functions are unique up to normalization. One can furthermore show that the two-point functions of $\frac{1}{2}$-BPS purely chiral objects vanish. For purely $\frac{1}{2}$-BPS operators, we find that the three point functions are also completely fixed by symmetry, up to a normalization identified with the structure constants. However, for the three-point correlation functions depending on three full points, an invariant superconformal cross-ratio appears, making them depend on an arbitrary function of the cross-ratio. Using superspace analyticity and Bose symmetry allows us for $\frac{1}{3}$-BPS operators to fix this function to be a polynomial of specific degree. For the stress-tensor multiplet, this gives the three-point correlation functions up to a number that we identify with the central charge.

\paragraph{Notation.}
We will denote by $1,2,\dots$ the chiral points $(\mathcal{X}_{+},W)_{1,2,\dots}$,
by $\bar{1},\bar{2},\dots$ the anti-chiral points $(\mathcal{X}_{-},\bar{W})_{\bar{1},\bar{2},\dots}$,
and by $\bf{1},\bf{2},\dots$ the full superspace points $(X,W,\bar{W})_{\bf{1},\bf{2},\dots}$.
We denote the $\tfrac{1}{2}$-BPS operators  of type $(3,B,+)$ and $(3,B,-)$ as $\varphi_k$ and $\bar{\varphi}_k$.
The $\tfrac{1}{3}$-BPS operators  of type $(3,B,2)$ will be denoted as $\mathcal{W}_k$ with the stress-tensor $\mathcal{T}\equiv \mathcal{W}_{k=1}$. 
\paragraph{Two-point structures.}
It is not hard to see that there are three basic $G_{>0}$ invariants that can be constructed out of two points
\beq\label{G>_inv}
\mathcal{X}_{1\bar{2}}\colonequals \mathcal{X}_{1,+}-\mathcal{X}_{2,-}-\bar{W}_2\wedge W_1\,,
\qquad
W_{12}\colonequals W_{1}-W_{2}\,,
\qquad
\bar{W}_{\bar{1}\bar{2}}\colonequals\bar{W}_{1}-\bar{W}_{2}\,.
\eeq
The matrix $\mathcal{X}_{1\bar{2}}$ is graded antisymmetric in the sense of \eqref{eq:defantisym}. The $G_{>0}$ invariant $\mathcal{X}_{2\bar{1}}$ is not independent, since we have the identity 
$\mathcal{X}_{1\bar{2}}+\mathcal{X}_{2\bar{1}}=\bar{W}_{\bar{1}\bar{2}}\wedge W_{12}$.
It follows from the transformations in Table \ref{tab:Xpmtransf}, that under $G_{(-,0)}$,  $G_{(0,-)}$
\beq
\label{eq:transformationpropertiesofthemathcalX}
\mathcal{X}_{1\bar{2}}\mapsto h_1^{}\,\mathcal{X}_{1\bar{2}}\,h_1^{st}\,,
\qquad
\mathcal{X}_{1\bar{2}}\mapsto \bar{h}_{\bar{2}}^{}\,\mathcal{X}_{1\bar{2}}\,\bar{h}_{\bar{2}}^{st}\,,
\eeq
where we have defined
$h_{\ell}\colonequals(\mathbbm{1}_4+W_{\ell} \,c)^{-1}$ and
$\bar{h}_{\bar{\ell}}\colonequals(\mathbbm{1}_4+\bar{W}_{\ell}\,\bar{c})^{-1}$.

There is a systematic way to produce covariant quantities in terms of coset coordinates.
This simple construction is sketched is Appendix \ref{App:covariants_fromcoset}.
 Applying this procedure to our case one obtains the covariant combination
\beq
\mathbf{X}_{\mathbf{1}\tilde{\mathbf{2}}}\,\colonequals \,
\mathcal{X}_{1\bar{2}}+W_{12}\,(\bar{W}_{\bar{1}\bar{2}})^{st}\,.
\eeq
Notice that we wrote this quantity in terms of the $G_{>0}$ combinations 
\eqref{G>_inv}.
As opposed to $\mathcal{X}_{1\bar{2}}$, the matrix $\mathbf{X}_{\mathbf{1}\tilde{\mathbf{2}}}$ is not graded antisymmetric, 
but satisfies the property 
$\Sigma\,\mathbf{X}_{\mathbf{1}\tilde{\mathbf{2}}}^{st}=\mathbf{X}_{\mathbf{2}\tilde{\mathbf{1}}}^{}$.
As explained in Appendix \ref{App:covariants_fromcoset}, its transformation properties under $OSP(6|4)$ take the form
\beq\label{X1tilde2_transf}
\mathbf{X}_{\mathbf{1}\tilde{\mathbf{2}}}\mapsto
H^{}_{\mathbf{2}}\,
\mathbf{X}_{\mathbf{1}\tilde{\mathbf{2}}}\,
H^{st}_{\mathbf{1}}\,,
\eeq
where $H_{\mathbf{\ell}}$ depends on the point $\mathbf{\ell}$ as well as on the group element $g$, see \eqref{Hdef:app} for its explicit form. 
The compatibility between the transformation properties \eqref{eq:transformationpropertiesofthemathcalX} and \eqref{X1tilde2_transf}  can be verified using the identity
\beq
\label{eq:greatX12identity}
\mathcal{X}^{}_{1\bar{2}} \,\mathbf{X}_{\mathbf{1}\tilde{\mathbf{2}}}^{-1}\,\mathcal{X}^{}_{2\bar{1}}=
\mathbf{X}^{}_{\mathbf{2}\tilde{\mathbf{1}}}\,.
\eeq
Before turning to the  discussion of two point functions, let us introduce some more building blocks.
We recall the definitions of  superdeterminant and super-Pfaffian\footnote{Due to our choice of gradation \eqref{eq:defPiandSigma}, the superdeterminant is defined to be the inverse of the one commonly found in the literature.} 
\beq
\text{sdet}
\begin{pmatrix}
A & B \\
C & D
\end{pmatrix}\,\colonequals \,
\frac{\det(D)}{\det(A-BD^{-1}C)}
\,=\,
\frac{\det(D-CA^{-1}B)}{\det(A)}\,,
\eeq
\beq\label{sPfdef}
\text{sPf}
\begin{pmatrix}
A & sB \\
B^t & D
\end{pmatrix}\,\colonequals \,
\frac{\text{Pfaff}(D-s B^tA^{-1}B)}{\sqrt{\det(A)}}\,=\,
\frac{\text{Pfaff}(D)}{\sqrt{\det(A-sB D^{-1}B^t)}}\,.
\eeq
Notice that while the definition of the superdeterminant makes sense for any square supermatrix, 
the definition of super-Pfaffian requires the matrix to be graded antisymmetric. 
The sign $s$ in  \eqref{sPfdef} corresponds to two type of graded antisymmetric matrices.
Specifically,   $s=1$, respectively $s=-1$, corresponds to the condition
$\mathcal{X}^{st}=-\Sigma\mathcal{X}$, respectively $\mathcal{X}^{st}=-\mathcal{X}\Sigma $.
In this work we will only need the ordinary Pfaffian of $2\times 2$ antisymmetric matrices which we define as
$\text{Pfaff}
\left(\begin{smallmatrix}0&t\\-t&0\end{smallmatrix}\right)=t$. 
It is then easy to check that $\text{Pfaff}(D^{-1})=-\text{Pfaff}(D)^{-1}$.
The super-Pfaffian satisfies the identity
\beq\label{sPFsdetidentity}
\text{sPf}(A\mathcal{X}A^{st})\,=\,\text{sdet}(A)\,\text{sPf}(\mathcal{X})\,.
\eeq
 From now on, we shall use the notation 
 \beq
 \label{eq:definitionbracketij}
 (i\bar{\jmath})\equiv \text{sPf}(\mathcal{X}_{i\bar{\jmath}})\,.
 \eeq
 It follows from \eqref{eq:transformationpropertiesofthemathcalX} and \eqref{sPFsdetidentity} that $ (i\bar{\jmath})$ transforms nicely, i.e. $ (i\bar{\jmath})\mapsto  \omega_i(g)\,\bar{\omega}_{\bar{\jmath}}(g) (i\bar{\jmath})$ with the $\omega_i$ and $\bar{\omega}_{\bar{\jmath}}$ taken from \eqref{eq:actionofomega}.

Using the identity \eqref{sPFsdetidentity}, the definition of the super-Pfaffian and rewriting \eqref{eq:greatX12identity} as
$\mathcal{X}_{2\bar{1}}=\mathbf{X}_{\mathbf{1}\tilde{\mathbf{2}}}\left(\mathcal{X}_{1\bar{2}}^{-1}\Sigma\right)\mathbf{X}_{\mathbf{1}\tilde{\mathbf{2}}}^{st}$, 
it is easy to see that 
\beq
\label{eq:superdeterminantdecomposition}
\text{sdet}(\mathbf{X}_{\mathbf{1}\tilde{\mathbf{2}}})=-(1\bar{2})(2\bar{1})\,.
\eeq
Hence, the only objects available to build the two-point functions are the super-Pfaffians \eqref{eq:definitionbracketij} and the matrix elements of $\mathbf{X}_{\mathbf{1}\tilde{\mathbf{2}}}$.

\paragraph{Two-point functions.}

It follows from the transformation properties of $(1\bar{2})$ and of $\mathbf{X}_{\mathbf{1}\tilde{\mathbf{2}}}$ that the two-point function of two generic operators \eqref{long_multi} takes the form
\beq\label{generaltwopoint}
\langle\mathcal{O}^{}_{\underline{\mathsf{A}}}(\mathbf{1})
\mathcal{O}^*_{\underline{\mathsf{B}}}(\mathbf{2})\rangle\,=\,
(1\bar{2})^{b_+}\,
(2\bar{1})^{b_-}\,
\mathcal{I}_{\underline{\mathsf{A}},\underline{\mathsf{B}}}(\mathbf{X}_{\mathbf{1}\tilde{\mathbf{2}}}^{-1})\,,
\eeq
where we remind that $*$ just exchanges the labels $b_+$ and $b_-$.
The superconformal covariance of  \eqref{generaltwopoint} is guaranteed if 
$\mathcal{I}_{\underline{\mathsf{A}},\underline{\mathsf{B}}}$ satisfies the identity
 \beq
\rho^{\underline{\mathsf{C}}}_{\underline{\mathsf{A}}}(M_{\mathbf{1}}(g))\,
 \mathcal{I}_{\underline{\mathsf{C}},\underline{\mathsf{D}}}(\mathbf{X}_{\mathbf{1}\tilde{\mathbf{2}}}^{-1})\,
 \rho^{\underline{\mathsf{D}}}_{\underline{\mathsf{B}}}(M_{\mathbf{2}}(g))\,=\,
 \mathcal{I}_{\underline{\mathsf{A}},\underline{\mathsf{B}}}(g\,\circ\,\mathbf{X}_{\mathbf{1}\tilde{\mathbf{2}}}^{-1})
 \eeq
 for $g\in OSP(6|4)$, where $M_{\mathbf{\ell}}(g)$ was introduced in \eqref{fromSOSPtoSU22}.
For $\frac{1}{2}$-BPS and $\frac{1}{3}$-BPS  operators, see Table~\ref{tab:nonlong reps} and equation \eqref{new_Osp_labels}, this two-point function reduces to 
\beq
\label{TwoPoint12BPS}
\langle\varphi_k(1)
\bar{\varphi}_k(\bar{2})\rangle\,=\,
(1\bar{2})^{k}\,,
\qquad
\langle\mathcal{W}^{}_{k}(\mathbf{1})
\mathcal{W}^{}_k(\mathbf{2})\rangle\,=\,
(1\bar{2})^{k}\,
(2\bar{1})^{k}\,.
\eeq
Setting the fermions to zero in the first part of \eqref{TwoPoint12BPS}, we obtain 
\beq
\label{eq:definitiony1bar2andtwopoints}
(1\bar{2})^{k}=\frac{y_{1\bar 2}^k}{|x_{12}|^k}=\frac{(V_1\cdot \bar{V}_2)^k}{|x_{12}|^k}\,,\qquad y_{1\bar 2}\colonequals y_{+,1}-y_{-,2}-\epsilon_{ab}v_1^a\bar{v}_2^b\,,
\eeq
with $y_{\pm}$ defined under \eqref{eq:oldrelationvwithyandomega}.
In particular, for $k=1$  we recover the two point correlation function $\langle\phi(x_1)
\phi(x_2)\rangle=\tfrac{1}{\sqrt{|x_{12}^2|}}$ of two free scalars in $d=3$.

\paragraph{Three-point structures.}
We define the basic three-point covariant $\Lambda_{1\bar{2},\mathbf{3}}$ as
\beq
\label{eq:definitionLambda123}
\Lambda_{1\bar 2,\mathbf{3}}\,:=\,
\mathcal{X}_{1\bar 3}^{-1}+\mathcal{X}_{3\bar 2}^{-1}
-\left(\mathcal{X}_{1\bar 3}^{-1}W_{13}\right)\wedge \left(\mathcal{X}_{3\bar 2}^{-1}\bar{W}_{\bar 2\bar 3}\right)\Sigma\,=\,
\mathcal{G}^{}_{1\bar{2},\mathbf{3}}\,
\mathcal{X}_{3\bar 2}^{-1}\,
\mathcal{X}_{1\bar 2}^{}\,
\mathcal{G}^{-1}_{1\bar{2},\mathbf{3}}\,
\mathcal{X}_{1\bar 3}^{-1}\,,
\eeq
where
$\mathcal{G}_{1\bar{2},\mathbf{3}}=
\mathbbm{1}+\,\mathcal{X}_{1\bar 3}^{-1}\,W^{}_{13}\,\bar{W}_{\bar{2}\bar{3}}^{st}$.
The matrix $\Lambda^{-1}_{1\bar 2\mathbf{3}}$ is graded anti-symmetric in the sense of \eqref{eq:defantisym}
and transforms as
\beq\label{Lambdatransf}
\Lambda_{1\bar 2,\mathbf{3}}\mapsto (h_3^{st})^{-1}\,
\Lambda_{1\bar 2,\mathbf{3}}(h_3)^{-1}\,,
\qquad \Lambda_{1\bar 2,\mathbf{3}}\mapsto (\bar{h}^{st}_{\bar 3})^{-1}\,
\Lambda_{1\bar 2,\mathbf{3}}(\bar{h}_{\bar 3})^{-1}\,,
\eeq
under $G_{(-,0)}$, respectively $G_{(0,-)}$ transformations.
 Its covariance under  $G_{(0,0)}$ and invariance under $G_{>}$ are manifest.
In order to find the covariant $\Lambda$ and prove its properties  it is convenient to 
use superconformal transformations to map the points $\{1,\bar 2,\mathbf{3}\}$ to either 
$\{(\Lambda^{-1},0),(\infty,0),(0,0,0)\}$
or
$\{(\infty,0),(\Lambda^{-1},0),(0,0,0)\}$.
This procedure is explained in detail in Appendix \ref{App:covariance_Lambda}.
Concerning the second expression for $\Lambda$ in \eqref{eq:definitionLambda123}, we observe that $\mathcal{G}_{1\bar{2},\mathbf{3}}$ intertwines
 $\mathcal{X}_{1\bar{2}}-\mathcal{X}_{3\bar{2}}$ with $\mathcal{X}_{1\bar{3}}$ as follows
\beq\label{Gintertw}
\mathcal{X}_{1\bar 2}-\mathcal{X}_{3\bar 2}=
\left(\Sigma\mathcal{G}_{1\bar{2},\mathbf{3}}^{st}\Sigma\right)\,
\mathcal{X}_{1\bar 3}\,
\mathcal{G}_{1\bar{2},\mathbf{3}}^{}\,.
\eeq
We observe that we have the following transformation properties for $\Lambda_{1\bar 2,\mathbf{3}}$ under exchange of the points:
\beq
\label{eq:Lambdatransformation}
\Lambda_{1\bar{3},\mathbf{2}}=\left(\mathbf{X}_{\mathbf{3}\tilde{\mathbf{2}}}\Lambda_{1\bar{2},\mathbf{3}}\mathbf{X}_{\mathbf{2}\tilde{\mathbf{3}}}\right)^{-1}\,,\qquad  \Lambda_{3\bar{2},\mathbf{1}}=\left(\mathbf{X}_{\mathbf{3}\tilde{\mathbf{1}}}\Lambda_{1\bar{2},\mathbf{3}}\mathbf{X}_{\mathbf{1}\tilde{\mathbf{3}}}\right)^{-1}\,.
\eeq
We can prove these identities by simply first observing that both sides of the equations transform the same way and then going in a frame where, in the first case $W_{12}=\bar{W}_{\bar 2\bar 3}=0$ and in the second $W_{13}=\bar{W}_{\bar 1\bar 2}=0$.

Given three full super-points, one can construct the  superconformal invariant 
\begin{equation}\label{threepointC}
 C_{\mathbf{123}}\colonequals\frac{(1\bar{3})(2\bar{1})(3\bar{2})}{(1\bar{2})(2\bar{3})(3\bar{1})}\,.
\end{equation}
The  invariance  of $C_{\mathbf{123}}$ can be shown using the transformation properties \eqref{eq:transformationpropertiesofthemathcalX} and  the identity \eqref{sPFsdetidentity}.
 Exchanging two points sends $C_{\mathbf{123}}$ to its inverse. 
The object $C_{\mathbf{123}}$ is the only invariant one can build out of three full points.
If the  Gra\ss mann variables are set to zero $C_{\mathbf{123}}$ depends only on the R-symmetry coordinates, 
which is compatible with the fact that there are no conformal cross-ratios made of three points.

To conclude this paragraph, let us connect the covariant $\Lambda_{1\bar{2},\mathbf{3}}$ to the invariant $C_{\mathbf{123}}$.
We can compute the super-Pfaffian of $\Lambda_{1\bar{2},\mathbf{3}}$ from which follows
\beq
\label{eq:superPfaffianLambda}
\text{sPf}(\Lambda_{1\bar 2,\mathbf{3}})=\frac{(1\bar 2)}{(1\bar 3)(3\bar 2)}\,\quad
\Longrightarrow
\quad
C_{\mathbf{123}}\,=\,\frac{\text{sPf}(\Lambda_{2\bar 1,\mathbf{3}})}{\text{sPf}(\Lambda_{1\bar 2,\mathbf{3}})}\,.
\eeq
The first equality in \eqref{eq:superPfaffianLambda} follows from the second form of $\Lambda$ in 
 \eqref{eq:definitionLambda123}. In addition, we can define the matrix $M_{\mathbf{1}\mathbf{2}\mathbf{3}}\colonequals \Lambda_{2\bar{1},\mathbf{3}}\Lambda_{1\bar{2},\mathbf{3}}^{-1}$. Thanks to \eqref{Lambdatransf} it transforms covariantly so that its eigenvalues are invariant. They are easily found to be $-1$ and $C_{\mathbf{123}}$, both with multiplicity two. Thanks to \eqref{eq:Lambdatransformation}, permuting the points in $M_{\mathbf{1}\mathbf{2}\mathbf{3}}$ sends the matrix to its inverse up to conjugation and hence does not give new invariants.

\paragraph{Three-point functions.} We will now discuss a few three-point functions  constructed using the covariant objects just introduced and imposing the requirement of superspace analyticity.
Let us list a few three-point functions of BPS operators ordered by  decreasing supersymmetry. First we note that the three-point function of only chiral or only anti-chiral operators vanishes; this is 
shown in Appendix \ref{App:chiral structures}.
For mixed correlators we have
\beq
\label{threePoint12BPS}
\langle\varphi_{k_1}(1)
\bar{\varphi}_{k_1+k_3}(\bar{2})\,
\varphi_{k_3}(3)
\rangle\,=\,
(1\bar{2})^{k_1}\,
(3\bar{2})^{k_3}\,,
\eeq
\beq
\label{threePoint12BPS13BPS}
\langle\varphi_{k}(1)
\bar{\varphi}_{k}(\bar{2})\,
\mathcal{W}_{\ell}(\mathbf{3})
\rangle\,=\,
(1\bar{2})^{k-\ell}\,
(1\bar{3})^{\ell}\,
(3\bar{2})^{\ell}\,.
\eeq
Superspace analyticity implies that 
 $\ell \leq k$ in \eqref{threePoint12BPS13BPS}. The same argument forces the correlation functions involving two $\tfrac{1}{3}-$BPS and one $\tfrac{1}{2}-$BPS operator to vanish. Generalizing, we find for three generic $\frac{1}{3}-$BPS operators
\beq
\begin{split}
\label{eq:threegenericthirdBPScorrelationfunction}
\langle\mathcal{W}_{k_1}(\mathbf{1})\,
\mathcal{W}_{k_2}(\mathbf{2})
\mathcal{W}_{k_3}(\mathbf{3})
\rangle\,=\,&
(1\bar{2})^{k_1}(2\bar{1})^{k_2-k_3}(2\bar{3})^{k_3}(3\bar{1})^{k_1-k_2+k_3}(3\bar{2})^{k_2-k_1}P_{K}(C_{\mathbf{1}\mathbf{2}\mathbf{3}})\,,
\end{split}
\eeq
where $P_{K}$ is an arbitrary polynomial of degree 
\beq
K\colonequals \min(k_1,k_3,k_1+k_3-k_2)-\max(0,k_1-k_2,k_3-k_2)\,,
\eeq
with the understanding that if $K<0$, then the correlation function \eqref{eq:threegenericthirdBPScorrelationfunction} vanishes. If two of the fields are identical, taking without loss of generality  $k_1=k_2=k$ and $k_3=\ell$, then the correlation function \eqref{eq:threegenericthirdBPScorrelationfunction} is further constrained by Bose symmetry, i.e.~invariance under the permutation of the first two operators. This permutation symmetry implies that the polynomial is constrained as 
 \beq\label{Pcondition}
 P_K(x)\,=\,x^\ell\,P_K(x^{-1})\,.
 \eeq
In particular, if all the fields are identical, we have
  \beq
\label{threePoint13BPS}
\langle\mathcal{W}_{k}(\mathbf{1})\,
\mathcal{W}_{k}(\mathbf{2})
\mathcal{W}_{k}(\mathbf{3})
\rangle\,=\,
(1\bar{2})^{k}\,
(2\bar{3})^{k}\,
(3\bar{1})^{k}\,
P_{k}(C_{\mathbf{1}\mathbf{2}\mathbf{3}})\, ,
\eeq
with $P_k(x)$ a polynomial of degree $k$ in $x$ obeying $P_k(x)=x^k\,P_k(x^{-1})$.
Perhaps surprisingly, this is the same condition as the requirement of parity invariance of the three-point function, see also  \cite{Buchbinder:2015qsa} for a similar observation.
In the case of the stress-tensor multiplet, corresponding to $k=1$, the condition 
\eqref{Pcondition} gives a unique solution up to normalization. The normalization can be identified 
with the central charge.

Finally, the covariant $\Lambda_{1\bar{2},\mathbf{3}}$ enters the three-point function of a chiral, anti-chiral, and an arbitrary operator with label $\{(b_+,b_-),(s,j,c_{2|2})\}$ as
\beq
\begin{split}
\label{threePoint12BPSANY}
\langle\varphi_{k+b_-}(1)
\bar{\varphi}_{k+b_+}(\bar{2})\,
\mathcal{O}_{\underline{\mathsf{A}}}(\mathbf{3})
\rangle&\,=\,
(1\bar{2})^{k}\,
(3\bar{2})^{b_+}\,
(1\bar{3})^{b_-}\,
t_{\underline{\mathsf{A}}}(\Lambda_{1\bar{2},\mathbf{3}})\,,
\end{split}
\eeq
where $t_{\underline{\mathsf{A}}}$ satisfies the identity
\beq
\rho_{\underline{\mathsf{A}}}^{\underline{\mathsf{B}}}(M_{\mathbf{3}}(g))\,
t_{\underline{\mathsf{B}}}(\Lambda_{1\bar{2},\mathbf{3}})\,=\,
t_{\underline{\mathsf{A}}}(g\,\circ\,\Lambda_{1\bar{2},\mathbf{3}})\,.
\eeq
 The class of operators for which  the three-point function \eqref{threePoint12BPSANY}
 is non-vanishing, is restricted by the requirement of superspace analyticity. A careful analysis of the constraints,
  as well as the analysis of more general three-point functions
   is left for future work.


\section{Four-point  functions of BPS operators}
\label{Sec:fourpoints}

In this section we wish to study the superconformal Ward identities for the correlation functions of BPS operators along the lines of
\cite{Dolan:2004mu}. 
The basic idea  is that, once we impose superconformal symmetry, the Ward identities follow from the requirement of 
harmonic superspace analyticity.
The setup in which the techniques of \cite{Dolan:2004mu} can be applied  without modification is described in the following.
The main condition is that we can send the  Gra\ss mann coordinates of the four points to zero by a superconformal transformation. This is the superspace version of the statement that the full four-point function is uniquely fixed in terms of the correlation function of the primary states.
The Ward identities are then the condition that the first term in the expansion of the four-point function in  the Gra\ss mann variables  is free from 
singularities in the harmonic R-symmetry coordinates.
Perhaps not surprisingly,  for our $3d$ $\mathcal{N}=6$ harmonic  superspace this setup is realized for the correlation functions of $\frac{1}{2}$-BPS operators of alternating chirality.
In this case, the Ward identities take the same form as the ones derived in \cite{Dolan:2004mu}  for  $\frac{1}{2}$-BPS supermultiplets in $\mathcal{N}=8$ theories in  three dimensions.

 The idea of superspace analyticity is very general and can be applied to situations beyond this setup. The main example is the case of four-point functions involving $\frac{1}{3}$-BPS as well as   $\frac{1}{2}$-BPS operators. The complications are due to the presence of nilpotent invariants.
 In the conclusions of the section we introduce a subspace of our harmonic superspace that we call super-line,
 as it is a supersymmetric extension of a line in spacetime.
The correlation functions simplify when restricted to this subspace and the Ward identities become more tractable\footnote{This type of  simplification is reminiscent to the example of conformal blocks in three dimensions.
Indeed, the latter are known in closed form only when the four points
are restricted to lie (up to conformal transformations) on a line, see~\cite{Rychkov:2015lca}.}.
Moreover, the notion of  super-line gives a convenient starting point
to discuss the cohomological reduction reviewed in the next section.

\subsection{Invariants and Ward identities}

The first step in the derivation of the Ward identities is the study of superconformal invariants made of four points,
generalizing the familiar conformal cross-ratios.
Let us consider  a group $G$ acting on a configuration space $\mathscr{C}$ and denote by 
 $S_{\underline{x}}$ the stability group of the configuration $\underline{x}\in\mathscr{C}$.
 Under mild assumptions, we have the following formula for 
 the number of invariants 
 \beq\label{numbinvformula}
 \text{Number of invariants}\,=\,
 \text{dim}\,\mathscr{C}-
  \text{dim}\,G
  +  \text{dim}\,S_{\underline{x}}\,.
 \eeq
 If we have fermionic symmetry acting on configuration spaces with Gra\ss mann coordinates, the situation is quite different.
 The reason for this is, roughly, that one cannot divide by nilpotent quantities, like the Gra\ss mann coordinates themselves.
 
\paragraph{The three-point invariant revisited.} Before turning to the discussion of four-point invariants, it is useful to 
revisit the derivation of the three-point invariant $C_{\mathbf{123}}$ given in \eqref{threepointC}.
A superconformal transformation allows us to map three full points $\{\mathbf{1},\mathbf{2},\mathbf{3}\}$ to 
\beq
\{\mathbf{1},\mathbf{2},\mathbf{3}\}\,\mapsto \,
\{(0,0,0),
(\Omega,
\left(\begin{smallmatrix}0\\u^a\end{smallmatrix}\right),
\left(\begin{smallmatrix}0\\\bar{u}^a\end{smallmatrix}\right)),(\infty, 0,0)\}\,,
\qquad
\Omega:=\begin{pmatrix}
\mathbbm{1}_2 & 0 \\
0 & \epsilon
\end{pmatrix}\,.
\label{Omegadef}
\eeq
This map comes as a result of the  composition of a $G_{>0}$ transformation that sends the first point to $(0,0,0)$, 
a $G_{<0}$ transformation that sends the third point to $(\infty, 0,0)$ and a $G_{(0,0)}$ transformation that sends 
the second point to the form above\footnote{This is strictly true  only when the two  vectors $u^a$ and $\bar{u}^a$
 are not proportional to each other. }. 
The class of frames given in \eqref{Omegadef} is acted upon by $SP(2)\times GL(1)$ as
$u^b\mapsto a\,S^b_c \,u^c $, $\bar{u}^b\mapsto a^{-1}\,S^b_c\, \bar{u}^c $ with $\det(S)=1$.
It is clear that  there is only one invariant  in this case, namely $\epsilon_{ab}\,u^a\,\bar{u}^b$.
In this frame, the  superconformal  invariant   $C_{\mathbf{123}}$ takes the form
\beq
C_{\mathbf{123}}\Big{|}_{\text{frame \eqref{Omegadef}}}\,=\,
\frac{1+\epsilon_{ab}\,u^a\,\bar{u}^b}{1-\epsilon_{ab}\,u^a\,\bar{u}^b}\,.
\eeq
We will now apply the same idea to various cases involving four points.

\paragraph{Superconformal invariants made of $\{1,\bar{2},3,\bar{4}\}$.}
We will now discuss superconformal invariants 
 constructed out of two chiral and two anti-chiral  points, taking without loss of generality the configuration $1,\bar{2},3,\bar{4}$.
In this case, the number of  Gra\ss mann coordinates of $1,\bar{2},3,\bar{4}$ is $4\times 6=24$ and is equal to the number of fermionic generators in $\mathfrak{osp}(6|4)$. This suggests that there are no nilpotent invariants in this case as we will show below.
It follows that  
all superconformal invariants made of points $1,\bar{2},3,\bar{4}$ are uniquely specified  by their \textit{body}, where we are using the terminology of \cite{Eden:1999kw}. In other words, they are the unique superconformal completion of conformal and R-symmetry invariants made 
of bosonic coordinates.

There is a systematic way to introduce such fully invariant objects.
They can be identified with the three distinguished eigenvalues $z,\bar{z},w$ of the supermatrix
\beq\label{Zfulldef}
\mathcal{Z}_{1\bar{2}3\bar{4}}:=\mathcal{X}_{1\bar{2}}^{}\,
\mathcal{N}\,
\mathcal{X}_{1\bar{4}}^{-1}\,
\mathcal{X}_{3\bar{4}}^{}\,
\mathcal{N}^{-1}\,
\mathcal{X}_{3\bar{2}}^{-1}\,,
\qquad
\mathcal{N}\,=\,\mathbbm{1}\, +\,
\left(\mathcal{X}_{1\bar{2}}-\mathcal{X}_{3\bar{2}}\right)^{-1}W^{}_{31}\,\bar{W}_{\bar{2}\bar{4}}^{st}\,.
\eeq
The body of this supermatrix is given by
\beq\label{Zbody}
\calZ_{1\bar 23\bar 4}\big{|}_{\text{ferm}=0}\,=\,\begin{pmatrix}
x_{12}^{}x_{14}^{-1}x_{34}^{}x_{32}^{-1} & 0 \\
0 & \frac{y_{1\bar{2}}\,y_{3\bar{4}}}{y_{1\bar{4}}\,y_{3\bar{2}}}\,\mathbbm{1}_2
\end{pmatrix}\,.
\eeq
where $y_{i\bar{\jmath}}$ was defined in \eqref{eq:definitiony1bar2andtwopoints}.
The fact that $z,\bar{z},w$ are invariants immediately follows from the identity
\beq\label{LamdaandZ}
\Lambda_{1\bar 4,\mathbf{3}}^{-1}\Lambda^{}_{1\bar 2,\mathbf{3}}\,=\,
\left(\mathcal{X}_{1\bar3}^{}\, \mathcal{G}_{1\bar{2},\mathbf{3}}\mathcal{X}_{1\bar2}^{-1}\right)
\calZ_{1\bar 23\bar 4}
\left(\mathcal{X}_{1\bar3}^{} \,\mathcal{G}_{1\bar{2},\mathbf{3}}\mathcal{X}_{1\bar2}^{-1}\right)^{-1}\,,
\eeq
with $\Lambda_{1\bar{2},\mathbf{3}}$ and $\mathcal{G}^{}_{1\bar{2},\mathbf{3}}$ defined in \eqref{eq:definitionLambda123} and subject to the transformation properties \eqref{Lambdatransf}.
See Appendix \eqref{App:covariance_monodromyNEW} for more details.
It is worth remarking that, while the matrix entries of \eqref{LamdaandZ} do not depend chirally on the point $\mathbf{3}$,
its eigenvalues do. 
Finally, we can show that the matrix $\calZ_{1\bar 23\bar 4}$ has only three distinguished eigenvalues
by going to the frame where all fermions are zero.
Using  $G_{>0}$ and $G_{<0}$ we can send the four points to  
\beq\label{framemapTEXT}
\Big{\{}1,\bar{2},3,\bar{4}
\Big{\}}\,
\mapsto\,
\Big{\{}(0,0),(\mathcal{Z}_-,\,0),(\mathcal{Z}_+,\,0),(\infty,0)
\Big{\}}\,,
\eeq
where $\mathcal{Z}_\pm$ are defined in \eqref{eq:definitionZpm} and
$ \mathcal{X}_{1\bar 2}^{-1}\, \calZ_{1\bar 23\bar 4}\, \mathcal{X}_{1\bar 2}^{}\,=\,
 (1-\mathcal{Z}_-^{-1}\mathcal{Z}_+)^{-1}$.
See Appendix \eqref{App:covariance_monodromyNEW} for further details.
It is not hard to see that the Gra\ss mann coordinates in $\mathcal{Z}_\pm$ can be set to zero by a $G_{(0,0)}$ transformation.
This implies that the eigenvalue structure of $\calZ_{1\bar 23\bar4}$ is the same as its body \eqref{Zbody} and  concludes the derivation.
For the later discussion, we notice that one 
 can  perform a $G_{(0,0)}$ transformation such that
\beq\label{FrameforZpm}
\mathcal{Z}_-\,\mapsto\,\Omega\,,
\qquad
\mathcal{Z}_+\,\mapsto\,\Omega\,\text{diag}(z_+,\bar{z}_+,w_+,w_+)\,,
\eeq
where $\Omega$ is given in \eqref{Omegadef}.
The frame \eqref{framemapTEXT} is left invariant by $SP(2)\times GL(1)\subset G_{(0,0)}$  transformations
corresponding to 
$A=\left(\begin{smallmatrix}\mathbbm{1}_2 &  0  \\0 & S\end{smallmatrix}\right)$ with $\det S=1$, $a\in \mathbb{C}^*$,
 see Table \ref{tab:Xpmtransf}. This is the, four-dimensional,  stability group of the four points $\{1,\bar{2},3,\bar{4}\}$.
The counting of invariants  agrees with the general formula  \eqref{numbinvformula} applied to conformal and 
R-symmetry respectively as
$2=12-10$ and $1=12-15+4$.  
Finally, let us notice that there is a $\mathbb{Z}_2$ symmetry acting on the frame \eqref{framemapTEXT},
\eqref{FrameforZpm} by exchanging the two eigenvalues $z_+,\bar{z}_+$.
As the true invariant is the set $\{z_+,\bar{z}_+\}$ rather then the individual eigenvalues, the
four-point correlation functions are required to be invariant under this discrete symmetry.
 Finally, we remark that due to the identification given below  \eqref{framemapTEXT}
  the eigenvalues $z$, $\bar{z}$ and $w$ are related to $z_+$, $\bar{z}_+$ and $w_+$ through
\beq
z=(1-z_+)^{-1}\,,\qquad  \bar{z}=(1-\bar{z}_+)^{-1}\,,\qquad w=(1-w_+)^{-1}\,.
\eeq

\paragraph{Superconformal invariants made of $\{\mathbf{1},\bar{2},3,\mathbf{4}\}$ and nilpotent invariants.}
In this case one can perform a superconformal transformation to map
\begin{align}\label{framemapTEXTtwofull1}
\Big{\{}\mathbf{1},\bar{2},3,\mathbf{4}
\Big{\}}\,
&\longmapsto\,
\Big{\{}(0,0,0),(\mathcal{Z}_-,\,
\bar{\mathcal{U}}_-)
,(\mathcal{Z}_+,\,
\mathcal{U}_+),(\infty,0,0)
\Big{\}}\,,\\
&\stackrel{G_{(0,0)}}{\longmapsto}\,
\Big{\{}(0,0,0),(\Omega,\,
\bar{\mathcal{U}}_-)
,(\Omega D,\,
\mathcal{U}_+),(\infty,0,0)
\Big{\}}\,,
\label{framemapTEXTtwofull2}
\end{align}
where $\Omega$ is given in \eqref{Omegadef}, 
$D:=\text{diag}(z_+,\bar{z}_+,w_+,w_+)$
and $\bar{\mathcal{U}}_-=\left(\begin{smallmatrix}\bar{\xi}^{\alpha}\\\bar{u}^a\end{smallmatrix}\right)$,
$\mathcal{U}_+=\left(\begin{smallmatrix}\xi^{\alpha}\\u^a\end{smallmatrix}\right)$.
The group $SP(2)\times GL(1)\times D_4$
 acts on the class of frames \eqref{framemapTEXTtwofull2}.
 It corresponds to a $G_{(0,0)}$ action with $a\in \mathbb{C}^*$ and
$A=\left(\begin{smallmatrix}\mathbf{g} &  0  \\0 & S\end{smallmatrix}\right)$ where $\det S=1$ and $\mathbf{g}$ is in the dihedral group $D_4$ generated by the two matrices $\mathbf{r}=\left(\begin{smallmatrix}  & -1 \\1 &  \end{smallmatrix}\right)$ and $\mathbf{s}=\left(\begin{smallmatrix}1 &   \\  & -1\end{smallmatrix}\right)$. Strictly speaking only $\mathbf{s}$ leaves the matrix $\Omega D$ invariant, since $\mathbf{r}$ permutes the two eigenvalues $z_+$ and $\bar{z}_+$. However, as we argued before, only the set of eigenvalues $\{z_+, \bar{z}_+\}$ is invariant, not the individual ones.

 After imposing $SP(2)$ symmetry, the generic invariant  under this action is given by a function of 
 $z_{+},\bar{z}_+,w_+,\epsilon_{ab}\,u^a\,\bar{u}^b,\xi^i,\bar{\xi}^i$ 
that is invariant with respect to the remaining $GL(1)\times D_4$. The $GL(1)$ part tells us that we have to combine $\xi^i$ with $\bar{\xi}^j$ and then the $ D_4$ removes the mixed combinations  $\xi^1\bar{\xi}^2$ and $\xi^2\bar{\xi}^1$. Hence, 
upon solving the constraints we arrive at the definition of the vector space
 \beq\label{Vdef}
 \mathbb{V}_{\text{inv}}:=
 \text{Span}_{\text{even}}\big{\{}1,
 \xi^1\bar{\xi}^1+\xi^2\bar{\xi}^2,
 \, \xi^1\xi^2  \bar{\xi}^1\bar{\xi}^2
 \big{\}}
 +
 \text{Span}_{\text{odd}}\big{\{}
 \xi^1\bar{\xi}^1-\xi^2\bar{\xi}^2
 \big{\}}\,,
 \eeq
 where $ \text{Span}_{\text{even/odd}}$ denotes the vector space over the field of functions of
 $z_{+},\bar{z}_+,w_+,\epsilon_{ab}\,u^a\,\bar{u}^b$ that are even/odd with respect to the exchange of $z_{+}$ 
 with $\bar{z}_+$.
 In this frame,
any invariant is an element  of $ \mathbb{V}_{\text{inv}}$ and vice versa. 
Notice that the stability group of $\{\mathbf{1},\bar{2},3,\mathbf{4}\}$ is one-dimensional\footnote{One can solve the stability conditions $S\,u=a^{-1}\,u$, $S\,\bar{u}=a^{+1}\,\bar{u}$, $\det{S}=1$ uniquely for $S$.}  so that 
the number of invariants for the bosonic subgroup agrees with  \eqref{numbinvformula} with $2=12-10$ and $2=16-15+1$.  
Let us further notice that number of Gra\ss mann variables in the frame \eqref{framemapTEXTtwofull2}
is equal to the number of Gra\ss mann coordinates of $\{\mathbf{1},\bar{2},3,\mathbf{4}\}$ minus the dimension of
 $\mathfrak{osp}(6|4)_{\bar{1}}$, namely $4=28-24$.

In order to discuss the superconformal Ward identity, it is necessary to present these invariants in 
the more general frame given by \eqref{framemapTEXTtwofull1}.
Apart from the three distinguished eigenvalues of the matrix $\mathcal{Z}:=\mathcal{Z}_+^{-1}\mathcal{Z}_-$, 
we have 
\begin{align}
\label{Ginvariantdef}
&g_1\colonequals \bar{\mathcal{U}}_-^{st}\,\mathcal{Z}_{-}^{-1}\,\mathcal{U}_+\,,& 
&g_2\colonequals\bar{\mathcal{U}}_-^{st}\,\mathcal{Z}_{+}^{-1}\,\mathcal{U}_+\,,&\nonumber\\
&g_3\colonequals\bar{\mathcal{U}}_-^{st}\,\mathcal{Z}_{+}^{-1}\,\mathcal{Z}_{-}^{}\mathcal{Z}_{+}^{-1}=\bar{\mathcal{U}}_-^{st}\,\mathcal{Z}\,\mathcal{Z}_{+}^{-1}\,\mathcal{U}_+\,,&
&g_4\colonequals\bar{\mathcal{U}}_-^{st}\,\mathcal{Z}_{-}^{-1}\,\mathcal{Z}_{+}^{}\mathcal{Z}_{-}^{-1}= \bar{\mathcal{U}}_-^{st}\,\mathcal{Z}^{-1}\,\mathcal{Z}_{-}^{-1}\,\mathcal{U}_+\,.&
\end{align}
These four invariants satisfy a single relation due to the  Cayley-Hamilton identity
 $\mathcal{Z}^3=c_1\mathcal{Z}^2+c_2 \mathcal{Z}+c_3$, with $c_i=c_i(z_+,\bar{z}_+,w_+)$, from which follows $\calZ \calZ_+^{-1}=c_1 \calZ_+^{-1}+c_2 \calZ_-^{-1}+c_3 \calZ^{-1} \calZ_-^{-1}$. Hence, we only take $\{g_{\ell}\}_{\ell=1}^3$ as independent.
Finding the precise change of basis that brings \eqref{Ginvariantdef} to \eqref{Vdef} can be done by writing the $g_{\ell}$ in the frame \eqref{framemapTEXTtwofull2}. We find  
\begin{align}
&g_1{|}_{\text{frame \eqref{framemapTEXTtwofull2}}}\,=\,\xi_1\bar{\xi}_1+\xi_2\bar{\xi}_2-\epsilon_{ab}u^a\bar{u}^b\,, \quad g_2{|}_{\text{frame \eqref{framemapTEXTtwofull2}}}\,=\,z_+^{-1}\xi_1\bar{\xi}_1+\bar{z}_{+}^{-1}\xi_2\bar{\xi}_2-w_+^{-1}\epsilon_{ab}u^a\bar{u}^b\,,\nonumber\\
&g_3{|}_{\text{frame \eqref{framemapTEXTtwofull2}}}\,=\,z_+\xi_1\bar{\xi}_1+\bar{z}_{+}\xi_2\bar{\xi}_2-w_+\epsilon_{ab}u^a\bar{u}^b\,,
\end{align}
so that clearly $g_i{|}_{\text{frame \eqref{framemapTEXTtwofull2}}}\in  \mathbb{V}_{\text{inv}}$.
We remark that the quartic element in \eqref{Vdef} can be obtained as a square of the quadratic one. Hence, the counting is complete and $g_1$, $g_2$ and $g_3$ together with the eigenvalues $z_+$, $\bar{z}_+$ and $w_+$ generate the full set of invariants.

\noindent
\emph{Remark:} One finds almost immediately two full invariants for the configuration $\{\mathbf{1},\bar{2},3,\mathbf{4}\}$, namely
\beq
\frac{(1\bar 2)(3\bar 4)}{(1\bar 4)(3\bar 2)}\,,\qquad \frac{(4\bar 2)(3\bar 1)}{(4\bar 1)(3\bar 2)}\,.
\eeq
Going to the frame \eqref{framemapTEXTtwofull2}, we can easily write them as cumbersome expressions involving the invariants $\{z_{+},\bar{z}_+,w_+,g_{\ell}\}$.

One may apply the same analysis to the  case of four points of the type 
$\{\mathbf{1},\bar{2},3,\bar{4}\}$. By going to a frame where these points take the value 
$\{(0,0,0),(\Omega,0),(\Omega D,\mathcal{U}),(\infty,0)\}$, it is not hard to see that the only invariants are the three 
distinguished entries of $D$. The counting agrees with  \eqref{numbinvformula} as the stability group is two-dimensional in this case. In particular there are no nilpotent invariants for this configuration.

\paragraph{Superconformal invariants made of $\{\mathbf{1},\mathbf{2},\mathbf{3},\mathbf{4}\}$.}
This case is similar to the previous one, so we will be brief. We can map four full points to the configuration
\beq\label{framemapTEXTALLfull}
\Big{\{}\mathbf{1},\mathbf{2},\mathbf{3},\mathbf{4}
\Big{\}}\,
\mapsto\,
\Big{\{}(0,0,0),(\Omega,\,
\mathcal{U},\bar{\mathcal{U}})
,(\Omega D,\,
\mathcal{V},\bar{\mathcal{V}}),
(\infty,0,0)\Big{\}}\,.
\eeq
In this case the stability group is trivial and the number of invariants of the bosonic subalgebra is given by $2=12-10$ 
and $5=20-15$
for conformal and R-symmetry respectively.  
Specifically, we have $\{z_+,\bar{z}_+\}$ for the conformal invariants and  five independent R-symmetry ones
which in this frame are given by 
\beq
\epsilon_{ab}u^a \bar{u}^b\,,\quad  \epsilon_{ab}v^a \bar{v}^b\,,\quad  \epsilon_{ab}u^a \bar{v}^b\,,\quad \epsilon_{ab}v^a \bar{u}^b\,,\quad \frac{\epsilon_{ab}u^a v^b}{\epsilon_{ab}\bar{u}^a \bar{v}^b}\,,
\eeq
where $\bar{\mathcal{U}}=\left(\begin{smallmatrix}\bar{\xi}^{\alpha}\\\bar{u}^a\end{smallmatrix}\right)$,
$\mathcal{U}=\left(\begin{smallmatrix}\xi^{\alpha}\\u^a\end{smallmatrix}\right)$, $\bar{\mathcal{V}}=\left(\begin{smallmatrix}\bar{\zeta}^{\alpha}\\\bar{v}^a\end{smallmatrix}\right)$,
and $\mathcal{V}=\left(\begin{smallmatrix}\zeta^{\alpha}\\v^a\end{smallmatrix}\right)$. As in the case of 
$\{\mathbf{1},\bar{2},3,\mathbf{4}\}$ there is a vector space parallel to 
the one in \eqref{Vdef}, which is the  $GL(1)\times D_4$ invariant subspace of the Fock space generated  by the
eight  Gra\ss mann variables $\xi, \bar{\xi}, \zeta$ and $\bar{\zeta}$.
We will denote a set of invariants that generate such space by $\{g_{\ell}\}$  in analogy with  \eqref{Ginvariantdef} .

\noindent
\emph{Remark:}
It follows from the  transformation property \eqref{X1tilde2_transf} that
\beq
\mathbf{M}_{\bf{1}\tilde{\bf{2}}\bf{3}\tilde{\bf{4}}}:=
\mathbf{X}_{\bf{1}\tilde{\bf{2}}}^{}\,
\mathbf{X}_{\bf{1}\tilde{\bf{4}}}^{-1}\,
\mathbf{X}_{\bf{3}\tilde{\bf{4}}}^{}\,
\mathbf{X}_{\bf{3}\tilde{\bf{2}}}^{-1}\,\,,
\qquad
\mathbf{M}_{\bf{1}\tilde{\bf{2}}\bf{3}\tilde{\bf{4}}}\mapsto
H_{\mathbf{2}}^{}\,
\mathbf{M}_{\bf{1}\tilde{\bf{2}}\bf{3}\tilde{\bf{4}}}\,
H_{\mathbf{2}}^{-1}\,.
\eeq
Hence, the four eigenvalues of the ``monodromy'' matrix $\mathbf{M}_{\bf{1}\tilde{\bf{2}}\bf{3}\tilde{\bf{4}}}$ are invariant. They can be written in terms of the full set of invariants for the four points.

\paragraph{Four-point  functions of $\frac{1}{2}$-BPS operators and their Ward identites.}
According to the discussion above, 
it follows from superconformal symmetry that
 the four-point function of $\frac{1}{2}$-BPS operators 
 of alternating chirality
 takes the form
\beq
\label{4points1_2BPS}
\langle
\varphi_k(1)\,
\bar{\varphi}_k(\bar{2})\,
\varphi_k(3)\,
\bar{\varphi}_k(\bar{4})
\rangle\,=\,
(1\bar{2})^k
(3\bar{4})^k
F_k(z,\bar{z},w)\,,
\eeq
where $z,\bar{z},w$ are the three distinguished eigenvalues of the super-matrix $\mathcal{Z}_{1\bar{2}3\bar{4}}$
given in \eqref{Zfulldef}.
While the expression \eqref{4points1_2BPS} possess the right  superconformal transformation properties for any 
$F_k(z,\bar{z},w)$, the requirement of superspace analyticity puts strong constraints on this function. 
In the following, we will spell out these constraints in the same spirit as \cite{Dolan:2004mu}.

The first condition on  $F_k(z,\bar{z},w)$ can be obtained by setting the  Gra\ss mann variables to zero.
In this case superspace, or harmonic, analyticity requires the four-point function \eqref{4points1_2BPS} 
to be a  polynomial in the R-symmetry coordinates. Due to the prefactor of \eqref{4points1_2BPS} and the form of the body of $w$ \eqref{Zbody}, it follows that  $F_k(z,\bar{z},w)$ is a polynomial of
degree $k$ in $w^{-1}$. 

The superconformal invariant variables $z,\bar{z},w$  admit an expansion in Gra\ss mann coordinates.
This expansion can be obtained efficiently by going to a frame where \eqref{Zfulldef} takes the form
\beq\label{ZwithFermions}
\calZ_{1\bar 23\bar 4}\,=\,
\begin{pmatrix}
z_0 & 0 & \Theta^{11} & \Theta^{12} \\
0& \bar{z}_0 &  \Theta^{21} & \Theta^{22} \\\
 \Theta^{12} & \Theta^{22} & w_0 & 0 \\
-  \Theta^{11} &- \Theta^{21}  & 0 & w_0
\end{pmatrix}\,.
\eeq
Notice that the lower left block is  $\epsilon\,\Theta^t$ since $\calZ_{1\bar 23\bar 4}$ is equal to the product of 
$\Omega$ with a graded antisymmetric matrix.
By computing the supertraces of powers of $\calZ_{1\bar 23\bar 4}$, we find that the eigenvalues of this matrix are
\beq
\label{eq:expanding eigenvalues in fermions}
\begin{split}
z&=z_0+\frac{2\,\Theta^{11}\Theta^{12}}{z_0-w_0}+\dots\,,
\\
\bar{z}&=\bar{z}_0+\frac{2\,\Theta^{21}\Theta^{22}}{\bar{z}_0-w_0}+\dots\,,\\
w&=w_0+\frac{\Theta^{11}\Theta^{12}}{z_0-w_0}+
\frac{\Theta^{21}\Theta^{22}}{\bar{z}_0-w_0}+\dots\,,
\end{split}
\eeq
where $\dots$ denote higher orders in fermions.
 These expressions have poles for $w_0=z_0$ and $w_0=\bar{z}_0$.
Inserting \eqref{eq:expanding eigenvalues in fermions} in $F_k(z,\bar{z},w)$, expanding in the fermions and requiring the absence of poles in the four-point function gives the superconformal Ward identities 
 \beq\label{eq:WardIdentities}
 \left(\partial_z+\tfrac{1}{2}\partial_w\right)F_k(z,\bar{z},w)\big{|}_{z=w}\,=\,0\,,\qquad
 \left(\partial_{\bar{z}}+\tfrac{1}{2}\partial_w\right)F_k(z,\bar{z},w)\big{|}_{\bar{z}=w}\,=\,0\,.
 \eeq
It is worth to remark that these conditions are
identical to the conditions derived for correlation functions of $\frac{1}{2}$-BPS
operators in $3d$ $\mathcal{N}=8$ theories in \cite{Dolan:2004mu}.
As opposed to the four-dimensional case, the relation \eqref{eq:WardIdentities}, cannot be solved explicitly.
If we write a power series expansion 
\beq
\label{eq:Fk as a polynomial}
F_k(z,\bar{z},w)=\sum_{n=0}^k\,
a_n(z,\bar{z})\,\left(\frac{\sqrt{z\bar{z}}}{w}\right)^n\,,
\eeq
 the Ward identities \eqref{eq:WardIdentities} 
give $k+1$ differential constraints on the coefficient functions $a_n(z,\bar{z})$.
These identities will play a crucial role in the derivation of superconformal blocks,
a problem that we leave for future work.
In the end of this section we will show how to solve the Ward identities in the case in which 
the operators lie on a super-line as defined in subsection \ref{subsec: super line} below.  
A simple solution of \eqref{eq:WardIdentities} is given by  \eqref{eq:Fk as a polynomial} with  $a_n(z,\bar{z})$ 
independent of $z,\bar{z}$. It is easy to verify that  free field theories belong to this class of solutions.

\paragraph{Mixed four-point  functions of $\frac{1}{2}$-BPS  and $\frac{1}{3}$-BPS  operators.}
From the discussion around equations \eqref{Vdef}  and \eqref{Ginvariantdef} for the configuration $\{\mathbf{1},\bar{2},3,\mathbf{4}\}$ we find that the general four-point correlation function of such operators takes the form
 \beq
\label{4points1_3BPS1_2BPS}
\begin{split}
\langle
\mathcal{W}_{p_1}(\mathbf{1})\,
\bar{\varphi}_k(\bar{2})\,
\varphi_k(3)\,
\mathcal{W}_{p_2}(\mathbf{4})
\rangle
\,=\,&(1\bar 4)^{p_1}(4\bar 1)^{p_1}(3\bar 2)^{k+p_1-p_2}(4\bar 2)^{p_2-p_1}(3\bar 4)^{p_2-p_1} \\&\times 
F_{k,p_1,p_2}(z,\bar{z},w,g_{1},g_2,g_3)\,,
\end{split}
\eeq
where the invariants  $\{g_{\ell}\}_{\ell=1}^3$ are defined in 
 \eqref{Ginvariantdef}. They are combinations of one bosonic invariant and two nilpotent ones.
 If the fermions are zero, then harmonic analyticity requires that $F_{k,p_1,p_2}(z,\bar{z},w,g_{\ell})
 $ is a polynomial explicitly given by 
 \beq
 \label{eq:expansionF2full for fermions zero}
F_{k,p_1,p_2}(z,\bar{z},w,g_{\ell})\big{|}_{\text{ferm}=0}\,=\,
 \sum_{\substack{m,n=\max(0,p_1-p_2)\\m+n\leq k+p_1-p_2}}^{p_1}a_{m,n}(z,\bar{z})\left(\frac{y_{1\bar 2}y_{3\bar 4}}{y_{1\bar 4}y_{3\bar 2}}\right)^m\left(\frac{y_{4\bar 2}y_{3\bar 1}}{y_{4\bar 1}y_{3\bar 2}}\right)^n\,,
 \eeq
 where $y_{i\bar{\jmath}}$ was defined in \eqref{eq:definitiony1bar2andtwopoints}.
 We remark that the R-symmetry combinations appearing in the polynomial expansion \eqref{eq:expansionF2full for fermions zero} can be related to the invariants $\{z,\bar{z},w,g_{\ell}\}$ 
 in \eqref{Vdef}, \eqref{Ginvariantdef} as
\beq
\frac{y_{1\bar 2}y_{3\bar 4}}{y_{1\bar 4}y_{3\bar 2}}{\Big|}_{\text{frame \eqref{framemapTEXTtwofull2}}}=
\frac{w}{1-w\,g_1}\Big{|}_{\text{ferm}=0}\,,
\qquad 
\frac{y_{4\bar 2}y_{3\bar1}}{y_{4\bar 1}y_{3\bar 2}}{\Big|}_{\text{frame \eqref{framemapTEXTtwofull2}}}=
\frac{1-w}{1-g_1\,w}\Big{|}_{\text{ferm}=0}\,.
\eeq
In order to derive the superconformal Ward identities in this case once should go to 
a frame analogue to \eqref{ZwithFermions},
expand the invariants evaluated in this frame in the Gra\ss mann coordinates $\Theta$'s and   
require cancellation of any spurious pole.
If the  invariants $g_{\ell}$ possess a regular expansion, as it seems to be the case, then the Ward identities will take the same form as in \eqref{eq:WardIdentities} with the extra variables $g_{\ell}$ playing the role of spectators.

\paragraph{Four-point  functions of $\frac{1}{3}$-BPS operators.}
It follows from superconformal symmetry that
 the four-point function of $\frac{1}{3}$-BPS operators
 takes the form
\beq
\label{4points1_3BPS}
\langle\mathcal{W}_k(\mathbf{1})\,
\mathcal{W}_k(\mathbf{2})\,
\mathcal{W}_k(\mathbf{3})\,
\mathcal{W}_k(\mathbf{4})\rangle
\,=\,
\left((1\bar{2})(2\bar{1})
(3\bar{4})(4\bar{3})\right)^k G_k(z,\bar{z},w, \{g_{\ell}\}
)\,,
\eeq
where $z,\bar{z},w$ are as in \eqref{4points1_2BPS} and $\{g_{\ell}\}$ are an appropriate generalization of the $\{g_{\ell}\}_{\ell=1}^3$ of \eqref{Ginvariantdef} that  generate the remaining invariants.
If the fermions are set to zero, as discussed below \ref{framemapTEXTALLfull},
 there are five independent R-symmetry invariants. We will denote them as $\{s_{R},\bar{s}_{R}\}$.
 On general grounds, we have that
 \beq
 \label{eq:Gforzerofermions}
 G_k(z,\bar{z},w, \{g_{\ell}\})\big{|}_{\text{ferm}=0}\,=\,\sum_{\alpha \,\in\, [k,0,k]\otimes [k,0,k]}
 b_{\alpha}(z,\bar{z})\,P_{\alpha}(\{s_{R},\bar{s}_{R}\})\,.
 \eeq
 The tensor product $[k,0,k]\otimes [k,0,k]$ is not multiplicity free, so that on group theory grounds 
 one would expect a higher number of structures in the sum above.
 However, as explained in detail in \cite{Rattazzi:2010yc},  this is not the case due to Bose symmetry.
 As in the case of three-point functions,  Bose symmetry implies parity symmetry for the correlator 
 \eqref{eq:Gforzerofermions}.
 It is still unclear whether there can exist terms in \eqref{4points1_3BPS},
  proportional to nilpotent invariants, that violate parity. 
The number of terms in this sum grows rather fast with $k$.
In the case of $k=1$, corresponding to the stress-tensor multiplet, the six terms entering the sum \eqref{eq:Gforzerofermions}
are given in Appendix \ref{app:R-invariants}.
We leave a systematic study on the analyticity constraints for higher $k$ to future work.

\subsection{Four-point function of operators on a super-line}
\label{subsec: super line}

We will now introduce the notion of a super-line.
This gives  a simplified setup to study solutions to the Ward identities 
and provides a bridge
with the cohomological construction reviewed in the next section.

\paragraph{Super-line: definition and symmetries.}
A non-null line in three-dimensional spacetime is fixed in terms of an invertible symmetric matrix $\gamma$ which we write as 
$\gamma^{\alpha\beta}=\tfrac{1}{2}(\kappa^{\alpha}\bar{\kappa}^{\beta}+\bar{\kappa}^{\alpha}\kappa^{\beta})$.
The line then corresponds to points of the form $x^{\alpha\beta}=t\,\gamma^{\alpha\beta}$ and is acted on by 
 $SL_{\text{line}}(2)$ transformations as $t\mapsto \tfrac{at+b}{ct+d}$.
A super-line is an extension of a non-null line in three dimensions  by fermionic and R-symmetry coordinates.
Points on a super-line transform under a subgroup of the superconformal group which extend the $SL_{\text{line}}(2)$.

We will define two types of super-lines and denote them respectively as
 $\mathscr{L}_{(2|2)}$ and $\mathscr{L}_{(4|4)}$,
where the subscript denotes the number of coordinates.
They are defined as subspaces of $\mathscr{M}_{(8|8)}$ for which the coordinates \eqref{eq:coordexpl}
take the form
\beq
X_{
\Upsilon
}
\,=\,
\begin{pmatrix}
t\,\gamma^{\alpha\beta} & 
-\tfrac{1}{2}\vartheta\,h^b\,\bar{\kappa}^{\alpha}-\tfrac{1}{2}\bar{\vartheta}\, \bar{h}^b\,\kappa^{\alpha}\\
-\tfrac{1}{2}\vartheta\,h^a\,\bar{\kappa}^{\beta}-\tfrac{1}{2}\bar{\vartheta}\,\bar{h}^a\,\kappa^{\beta}
& \tfrac{1}{2}y\,\epsilon^{ab}
\end{pmatrix}\,,
\eeq
where $\Upsilon:=\left(\begin{smallmatrix}t &\vartheta\\ 
\bar{\vartheta} &y \end{smallmatrix}\right)$. 
The two-component objects
  $h_a$, $\bar{h}_a$ serve to specify the super-line in addition to 
$\kappa^{\alpha}$ and $\bar{\kappa}^{\alpha}$.
Moreover we use the normalizations
$\bar{\kappa}^{\alpha}\epsilon_{\alpha\beta} \kappa^{\beta}=1$, $\bar{h}_a \epsilon^{a b} h_b = 1$, and raise and lower 
indices using $\epsilon$. 
The coordinates $W,\bar{W}$ are set to zero for the super-line $\mathscr{L}_{(2|2)}$, 
while they are given by
\beq
W_{\omega}=\begin{pmatrix}
\kappa^{\alpha}\xi \\
h^{a}v
\end{pmatrix}\,,
\qquad
\bar{W}_{\bar{\omega}}=\begin{pmatrix}
\bar{\kappa}^{\alpha}\bar{\xi} \\
\bar{h}^{a}\bar{v}
\end{pmatrix}\,,
\eeq
where $\omega:=\left(\begin{smallmatrix}\xi\\ 
v \end{smallmatrix}\right)$ and
$\bar{\omega}:=\left(\begin{smallmatrix}\bar{\xi}\\ 
\bar{v} \end{smallmatrix}\right)$,
for the $\mathscr{L}_{(4|4)}$ super-line. 
Both super-lines are acted on by the group $GL_{\text{line}}(2|2)\subset OSP(4|4)\subset OSP(6|4)$
 in the following way
\beq\label{SL22onUpsilon}
\Upsilon
\mapsto
\left(P\,\Upsilon+Q\right)
\left(R\,\Upsilon+S\right)^{-1}\,,
\qquad
\begin{pmatrix}P &Q\\ 
R&S \end{pmatrix}\,\in\,GL_{\text{line}}(2|2)\,,
\eeq
\beq
\omega\,\mapsto\,\left(R\,\Upsilon_++S\right)^{-1}\,\omega\,,
\qquad
\bar{\omega}\,\mapsto \left(P'+\Upsilon_-R'\right)^{-1}\,\bar{\omega}\,,
\eeq
where  $\left(\begin{smallmatrix}P' &Q'\\ 
R'&S '\end{smallmatrix}\right):=\left(\begin{smallmatrix}P &Q\\ 
R& S\end{smallmatrix}\right)^{-1}$ and the grading  is understood from the form of $\Upsilon$.
Two remarks are in order. Firstly both definitions are 
compatible with  chirality constraints, for this reason the matrices $\Upsilon_{\pm}$ are well defined.
Secondly, the super-line $\mathscr{L}_{(2|2)}$ can be defined within the $\mathcal{N}=4$  superspace described 
in the end of Section \ref{sec: a novel superspace}. 
Using this observation it is not hard to see how  $GL_{\text{line}}(2|2)$ is embedded in the $\mathcal{N}=6$ 
superconformal group.
The super-line $\mathscr{L}_{(4|4)}$ is acted on by a larger group  which is isomorphic to $GL(2|3)$.
The extra generators sit in $G_{(0,\pm)}$, $G_{(\pm,0)}$.

\paragraph{Reduced Ward identities.}
The first step consists in restricting the superconformal invariants discussed at the beginning of this section to the super-line.
The $g_{\ell}$ invariants entering the four-point function involving at least two 
$\frac{1}{3}$-BPS operators vanish when restricted to the  $\mathscr{L}_{(2|2)}$ super-line.
The three types of four-point functions  studied above thus  collapse to the case of $\frac{1}{2}$-BPS operators.
This fact is easily explained  by recalling that the $\mathscr{L}_{(2|2)}$ super-line is defined within the $\mathcal{N}=4$
restriction of our $\mathcal{N}=6$ harmonic superspace described 
in the end of Section \ref{sec: a novel superspace}. 
 Let us describe how to derive and solve the Ward identities in this case.
 We have that
 \beq
\label{4points1_2BPSRED}
\frac{\langle
\varphi_k(1)\,
\bar{\varphi}_k(\bar{2})\,
\varphi_k(3)\,
\bar{\varphi}_k(\bar{4})
\rangle}{(1\bar{2})^k
(3\bar{4})^k
}\Big{|}_{\mathscr{L}_{(2|2)}}\,=\,
f_k(z,w)\,.
\eeq
In this case $z,w$ are the eigenvalues of the $(1|1)\times (1|1)$ supermatrix
$\left(\begin{smallmatrix}z_0 &\vartheta\\ 
\bar{\vartheta} &w_0 \end{smallmatrix}\right)$ and are
 given by $z_{0}+\tfrac{\vartheta \bar{\vartheta}}{z_{0}-w_0}$,
 $w_{0}+\tfrac{\vartheta \bar{\vartheta}}{z_{0}-w_0}$.
 Cancellation of the spurious pole for $z_0=w_0$ in the correlation function gives the Ward identities
 \beq\label{WI12BPSreduced}
  \left(\partial_z+\partial_w\right)f_k(z,w)\big{|}_{z=w}\,=\,0\,.
 \eeq
This equation coincides with the Ward identities \eqref{eq:WardIdentities}
 restricted to $z=\bar{z}$.
 The equation \eqref{WI12BPSreduced} is easy to solve. Moreover,
 it is easy to see that it implies that the function $f_{k}(z,z)$ is a piecewise constant.  The same derivation works for the correlators 
 once restricted to the super-line $\mathscr{L}_{(2|2)}$.
 The super-line $\mathscr{L}_{(4|4)}$ carries more structure as it truly distinguishes between $\frac{1}{2}$-BPS
 and $\frac{1}{3}$-BPS $\mathcal{N}=6$ operators.
 We leave the study of the Ward identities truncated to the $\mathscr{L}_{(4|4)}$ super-line for the future.

\section{Cohomological construction}
\label{sec:cohomological construction}

In this section we will review the cohomological construction of \cite{Chester:2014mea,Beem:2013sza}  in order to explain why the function $F_k(z,\bar{z},w)$ entering \eqref{4points1_2BPS} 
 collapses to a constant when  $(\bar{z},w) \to (z,z)$. 
The key observation made in 
\cite{Beem:2013sza} is that any superconformal theory whose
superconformal algebra possesses an $\mathfrak{sl}(2|2)$ subalgebra, 
admits a cohomological reduction to a simpler, possibly solvable, subsector.
This construction applied to the
 three-dimensional case 
  implies that any superconformal theory with $\Nm \geq 4$ contains a subsector of operators whose correlation functions are described by a $1d$ topological theory \cite{Chester:2014mea,LR}.

\paragraph{General Idea.}
Let us consider correlation functions
$\langle\Om_1(\mathsf{p}_1)\dots \Om_n(\mathsf{p}_n)\rangle$
 possessing a certain, possibly conformal, supersymmetry $\mathsf{Sym}$
and  
 identify a nilpotent fermionic generator $\mathbb{Q}\in \mathsf{Sym}$ and 
 the $\mathbb{Q}$-exact subalgebra $\tilde{\mathfrak{t}}\subset \mathsf{Sym}$.
 It then follows from a standard argument that under the condition 
  \beq\label{QkillsOperators}
 [\mathbb{Q},\,\Om_{\ell}(\mathsf{p}_{\ell})]\,=\,0\,,
 \qquad
\ell=1,\dots,n\,,
 \eeq
we can move each operator using the symmetry generated by $\tilde{\mathfrak{t}}$ without changing the value of the correlator.
This is the basic idea  used to define the chiral ring  in $\mathcal{N}=1$ four dimensional theories, see e.g.~\cite{Cachazo:2002ry}.
In    \cite{Beem:2013sza,Beem:2014kka}, 
this construction was  applied to the case in which $\mathsf{Sym}$ is the superconformal algebra  in order to identify a
solvable truncation of the conformal bootstrap equations in supersymmetric theories.

Let us now describe the  main features of the superconformal case.
Firstly, $\mathbb{Q}$ is a combination of ordinary supersymmetries and special-conformal supersymmetries. 
Secondly, the $\mathbb{Q}$-exact subalgebra $\tilde{\mathfrak{t}}$ is  a twisted version of spacetime symmetry, i.e.~a combination 
of conformal  and  R-symmetry generators.
Finally, in order to satisfy the condition \eqref{QkillsOperators}, not only do the operators $\Om_{\ell}$ have to be of a special type, 
but their positions $\mathsf{p}_{\ell}$  cannot be generic either.
The generators of $\mathsf{Sym}$ in the $\mathbb{Q}$ cohomology form the residual symmetry of the twisted correlators.

In all the cases analyzed  so far  the starting point is the identification of an  $\mathfrak{sl}(2|2)$ subalgebra of the superconformal algebra.
Once this is done, the rank one strange Lie superalgebra $\mathfrak{q}(1)\subset  \mathfrak{sl}(2|2)$ gives the desired structure:
$\tilde{\mathfrak{t}}$ is taken to be the  bosonic subalgera $\mathfrak{q}(1)_{\bar{0}} \simeq  \widetilde{\mathfrak{sl}(2)}$,
 $\mathbb{Q}$ is identified with the $ \widetilde{\mathfrak{sl}(2)}$-singlet generator in $\mathfrak{q}(1)_{\bar{1}}$, while the rest of  $\mathfrak{q}(1)_{\bar{1}}$ generates $\mathfrak{q}(1)_{\bar{0}}$ by commutation with $\mathbb{Q}$, thus making it 
 $\mathbb{Q}$-exact.
\paragraph{The three-dimensional case.}
Let us explain in more details the three-dimensional case analyzed in  \cite{Chester:2014mea}, \cite{LR}.
The relevant $\mathfrak{sl}(2|2)\subset \mathfrak{osp}(4|4)$ has been identified in Section \ref{subsec: super line}.
Applying the procedure of \cite{Beem:2013sza} outlined above  one obtains 
the nilpotent supercharge $\mathbb{Q}$ 
\be 
\mathbb{Q} = (\bar{h}_a \kappa^{\alpha} + h_a \bar{\kappa}^{\alpha}) \mathfrak{Q}^a_{\alpha}-(h^a \bar{\kappa}_{\alpha} - \bar{h}^a \kappa_{\alpha} )\mathfrak{S}_{a}^{\alpha}\,,
\ee
and the twisted $\widetilde{\mathfrak{sl}(2)}$ generated by
\beq
\mathsf{E}:=\,\gamma^{\alpha\beta}\,\mathfrak{P}_{\alpha\beta} + \mathfrak{r}\,,
\qquad
\mathsf{F}:=\,\gamma^{-1}_{\alpha\beta}\,\mathfrak{K}^{\alpha\beta} - \mathfrak{t}\,,
\qquad
\mathsf{H}:=
\mathfrak{D} - \mathfrak{U}\,,
\eeq
see Appendix \eqref{app: Algebra conventions from oscillator realization } for a summary of our conventions. 
Here $\{\gamma^{\alpha\beta}\,\mathfrak{P}_{\alpha\beta}, \gamma^{-1}_{\alpha\beta}\,\mathfrak{K}^{\alpha\beta}, \mathfrak{D} \}$ are the $\mathfrak{sl}_t(2)$ generators that exponentiate to  M\"obius transformations for the point on the line $t\mapsto \tfrac{at+b}{ct+d}$. The $\{\mathfrak{r}, -\mathfrak{t}, -\mathfrak{U}\}$ generators correspond to the $\mathfrak{sl}_y(2)$ factor and they exponentiate to $y\mapsto \tfrac{ay+b}{cy+d}$, compare to \eqref{SL22onUpsilon}.

The next step is to classify the cohomology of $\mathbb{Q}$ at the level of representations.
If an operator $\Om(0)$ at the origin is in the cohomology of $\mathbb{Q}$, the twisted translated operator $\widetilde{\Om}(\tau)$ obtained by translation with $\widetilde{\mathfrak{sl}(2)}$ is also in the cohomology, and $\Om(0)$ defines its cohomology class.
It is then easy to prove, using the argument reviewed above, 
that correlation functions of twisted operators
\be 
\langle \widetilde{\Om}(\tau_1) \cdots
 \widetilde{\Om}(\tau_n) \rangle = f(\tau_1,\cdots ,\tau_n)\, ,
\ee
 only depend on the ordering of the coordinates
and are therefore described by a $1d$ topological theory. 
In superspace language, a twisted translated operator $\widetilde{\Om}(\tau)$
corresponds to a  superfield, in a certain class  given below, restricted to a one-dimensional subspace of 
the super-line $\mathscr{L}_{(2|2)}$ with coordinate $\Upsilon=\tau\,\mathbbm{1}_2$.
We denote this subspace as $\mathscr{L}_{\tau}$.

It was shown in \cite{Chester:2014mea} that operators that belong to the cohomology of $\mathbb{Q}$ are also highest weights of $(2,B,+)$ multiplets, see Table \ref{tab:nonlong reps osp4}.
 The $(3,B,\pm)$ and $(3,B,2)$ multiplets studied in Section \ref{Sec:fourpoints} do contain a unique $(2,B,+)$ factor
 in their $\mathfrak{osp}(6|4) \to \mathfrak{osp}(4|4)$ decomposition:
\beq
  \begin{split}
(3,B,+) & \to \cdots + (2,B,+) + \cdots
\\
(3,B,-) & \to \cdots + (2,B,+) + \cdots
\\
(3,B,\,2\,) & \to \cdots + (2,B,+) + \cdots
\end{split}
\eeq
This can be verified using the superconformal characters of  \cite{Dolan:2008vc},
or by a superspace analysis as in the end of Section \ref{sec: a novel superspace}.
Hence,  both $\frac{1}{2}$-BPS and $\frac{1}{3}$-BPS $\mathcal{N}=6$ operators have a non-trivial representative in the 
$\mathbb{Q}$-cohomology.
As promised, the cohomological construction explains why $F_{k}(z,z,z)$, obtained by solving the 
Ward identities \eqref{eq:WardIdentities}, is a piecewise constant. 
The same reasoning can be applied without modification to the cases \eqref{4points1_3BPS1_2BPS} and 
\eqref{4points1_3BPS}. When restricted to the one dimensional subspace $\mathscr{L}_{\tau}$, these correlation functions are a priory functions of $z$, since $z=\bar{z}=w$ and $g_{\ell}=0$.
It follows from the cohomological construction, or equivalently from reduced Ward identities,
that such functions are piecewise constants.

\section{Conclusions and outlook}
\label{sec: conclusions}

In this article we introduced a new harmonic superspace suitable for performing calculations involving BPS operators in $3d$ $\calN=6$ SCFTs. We analyzed many structures appearing in correlators involving two, three, and four points: invariants, covariants, cross-ratios, and nilpotent quantities. Thanks to these results we obtained a collection of two- and three-point functions and gave partial results for the four-point functions of BPS multiplets.

As stated in the introduction, one of our main motivations in undertaking this task is to prepare the terrain for the $3d$ $\calN=6$ superconformal bootstrap. For the $\tfrac{1}{3}$-BPS stress-tensor multiplet work remains to be done, in particular regarding the nilpotent invariant structures that can appear in the correlator (see \cite{Eden:1999gh} for a discussion of nilpotent invariants in $\calN=4$ SYM). However, for $\tfrac{1}{2}$-BPS multiplets we succeeded in deriving the Ward identities and the stage is set to implement the bootstrap equations. One way to solve \eqref{eq:WardIdentities} is to write the correlator as a sum of superconformal blocks. Each superconformal block is then written as a sum of bosonic blocks with arbitrary coefficients, the Ward identities will imply relations among these coefficients which can be understood as a consequence of supersymmetry: different conformal families are connected by the action of supercharges inside a superconformal multiplet. Once this is done the crossing equations will be ready to be analyzed using the numerical techniques of \cite{Rattazzi:2008pe}. The bootstrap is of particular interest for $\tfrac{1}{2}$-BPS multiplets because monopole operators sit in this type of multiplet. Monopole operators are local but cannot be written in terms of the fundamental fields of the theory, and therefore the numerical bootstrap is perfect for their study. We plan to come back to this problem in the future.

As opposed to their four-dimensional $\mathcal{N}=4$ relatives, the three-point functions of $\frac{1}{3}$-BPS
operators in  ABJM depend non-trivially on the couplings of the theory.
The first non-trivial contributions in perturbation theory have been computed in 
 \cite{Young:2014lka,Young:2014sia}.
 It seems plausible to expect that these structure constants can be computed by localization.
 For the simplest  $\frac{1}{3}$-BPS operator, namely the stress-tensor supermultiplet,  the relevant  building blocks can be found in the  literature
  \cite{Jafferis:2010un,Closset:2012vg,Chester:2014fya}.
  One can speculate that, in the planar limit, this three-point structure is directly related to the function 
 $h(\lambda)$
 computed in \cite{Gromov:2014eha} by a combination of  localization and  integrability methods.

Another interesting direction to be explored is the three-dimensional analogue of the triality between  
correlation functions, scattering amplitudes, and Wilson loops along the lines of \cite{Alday:2010zy}.
As pointed out there, the duality between  correlation functions and Wilson loops in gauge theories
holds under quite general assumptions and in various dimensions.
On the other hand, the duality between scattering amplitudes and Wilson loops in planar ABJM is still 
not well understood. 
It is expected to hold as a consequence of integrability and passed non-trivial perturbative tests,
see  \cite{Henn:2010ps, Chen:2011vv, Bianchi:2011dg, Bianchi:2011rn, Bianchi:2014iia}. 
We expect that the correlation functions of the stress-tensor superfield in the harmonic superspace introduced in this paper will provide a bridge between these two objects and close the triangle.


\section*{Acknowledgments}
We thank M. Bianchi, B. Eden, E. Ivanov, E. Sokatchev
and W. Siegel  for useful discussions and feedback. 
P.L. and V.M. acknowledge support by the SFB 647 ``Raum-Zeit-Materie. Analytische und Geometrische Strukturen''. V.M. is also supported by the PRISMA cluster of excellence at the Johannes Gutenberg University in Mainz.
The authors would like to thank the Simons Summer Workshop 2015 where part of this work was performed.

\appendix
\addtocontents{toc}{\protect\setcounter{tocdepth}{1}}
\section{Conventions}
\label{App:covent}

For convenience, we would like to spell out in Table \ref{tab:TN=6} the decomposition of the stress-tensor multiplet in representations of the bosonic subalgebra.
This multiplet can be found in 
\cite{Nilsson:1984bj}, see also\footnote{References \cite{Flato:1984du,Andrianopoli:2008ea} seem to contain some mistake.} \cite{Papathanasiou:2009en}

\begin{table}
\centering
\renewcommand{\arraystretch}{1.4}
\begin{tabular}{%
|l    |>{\centering }m{3.7cm}
                |>{\centering }m{3.7cm}
             |>{\centering\arraybackslash}m{3.7cm}|
}
\hline
&
$(r_1,r_2,r_3)$
&$[p,q,r]$
&$\big(\Delta,s\big)$
\\\hline
&
$(1,1,0)$
&
$[1,0,1]$
&$\big(1,0^+\big)$
\\\hline
 &
$(1,0,0)$
\linebreak 
$(1,1,1)$
\linebreak 
$(1,1,-1)$
&
$[0,1,0]$
\linebreak 
$[2,0,0]$
\linebreak
$[0,0,2]$
&
$\big(\frac{3}{2},\frac{1}{2}\big)$
\\\hline
&
$(1,1,0)$
\linebreak 
$(0,0,0)$
&
$[1,0,1]$
\linebreak 
$[0,0,0]$
&
$\big(2,1\big)$
\\\hline
 &
$(1,1,0)$
&
$[1,0,1]$
&
$\big(2,0^-\big)$
\\\hline
 &
$(1,0,0)$
&
$[0,1,0]$
&
$\big(\frac{5}{2},\frac{3}{2}\big)$
\\\hline
&
$(0,0,0)$
&
$[0,0,0]$
&
$\big(3,2\big)$
\\\hline
\end{tabular}
\renewcommand{\arraystretch}{1.0}
\caption{
Decomposition of the $\mathcal{N}=6$ stress tensor supermultiplet
  $\mathcal{T}$ into representations of the 
bosonic subalgebra $\mathfrak{so}(6)\oplus\mathfrak{sp}(4)\subset \mathfrak{osp}(6|4)$. 
Finite dimensional 
representations of $\mathfrak{so}(6)$ are characterized by the
Dynkin labels  $(r_1,r_2,r_3)$. They can be converted to  $\mathfrak{su}(4)$ Dynkin
labels as  $(r_1,r_2,r_3)\mapsto[r_2+r_3,r_1-r_2,r_2-r_3]$.
Unitary irreducible representations of $\mathfrak{sp}(4,\mathbb{R})$ are uniquely specified by the pair $(\Delta,s)$.
The symbols  $0^{+}$, $0^-$ denote Lorentz scalars and pseudo-scalar respectively, i.e.~$\pm$ characterizes the transformation properties 
of the multiplet under parity. }
\label{tab:TN=6}
\end{table}

\begin{table}
\centering
\renewcommand{\arraystretch}{1.6}
\begin{tabular}{%
| l
                |>{\centering }m{3.1cm}
 |>{\centering }m{3.1cm}
             |>{\centering\arraybackslash}m{5.5cm}|
}
\hline
Type 
&Name 
& Fraction of $\mathfrak{Q}$-susy that kills the h.w.s.
&$\{(j_L,j_R),(\Delta,s)\}$
\\\hline
Short
& 
$(2,B,1)$ \linebreak 
$(2,B,+)$
\linebreak 
$(2,B,-)$

&
$1/4$ \linebreak 
$1/2$ \linebreak
$1/2$
&
$\{(k,k),(k,0)\}$ \linebreak 
$\{(\tfrac{1}{2}k,0),(\tfrac{1}{2}k,0)\}$ \linebreak
$\{(0,\tfrac{1}{2}k),(\tfrac{1}{2}k,0)\}$
\\\hline
Semi-Short
& 
$(2,A,1)$ \linebreak 
$(2,A,+)$ \linebreak
$(2,A,-)$
&
$1/8$ \linebreak 
$1/4$ \linebreak
$1/4$
&
$\{(k,k),(k+s+1,s)\}$ \linebreak 
$\{(k,0),(k+s+1,s)\}$ \linebreak
$\{(0,\tfrac{1}{2}k),(\tfrac{1}{2}k+s+1,s)\}$
\\\hline
Cons. Currents
& 
$(2,\,\text{cons})$
&
3/8
&
$\{(0,0),(s+1,s)\}$
\\\hline
\end{tabular}
\renewcommand{\arraystretch}{1.0}
\caption{Non-long representations of $\mathfrak{osp}(4|4)$ in the notation of 
 \cite{Dolan:2008vc}, see also table 1 in \cite{Chester:2014mea}.
 The first label is fixed as $2=\mathcal{N}/2$.
As opposed to the $\mathcal{N}=6$ case, for $\mathcal{N}=4$  the stress-tensor supermultiplet is of type $(2,\text{cons})$ for $s=0$.
This implies that it is not an isolated short operator and thus can recombine with other short operator to form a long multiplet.}
\label{tab:nonlong reps osp4}
\end{table}

\subsection{Algebra conventions from oscillator realization}
\label{app: Algebra conventions from oscillator realization }

There is an efficient way to present our conventions for the generators and the commutation relations of $\mathfrak{osp}(6|4)$.
It is based on the oscillator realization of $\mathfrak{osp}(6|4)$ that we will now review.
We introduce the the Heisenberg superalgebra $\mathcal{H}_{2|3}$ generated by the bosonic $\bar{\mathbf{a}}_{\alpha}, \mathbf{a}^{\alpha}$ and 
 the fermionic $\bar{\mathbf{c}}_i, \mathbf{c}^i$ oscillators with $\alpha=1,2$, $i=1,2,3$ subject to the relations
\beq
[\mathbf{a}^\alpha,\bar{\mathbf{a}}_{\beta}]\,=\,\delta^{\alpha}_{\beta}\,,
\qquad
\{\mathbf{c}^i,\bar{\mathbf{c}}_{j}\}\,=\,\delta_{j}^{i}\,.
\eeq
All the other graded commutators vanish. In the following we will denote $\mathbf{c}^3,\bar{\mathbf{c}}_3$ by 
$\mathbf{c},\bar{\mathbf{c}}$.
We define the representation $\rho: \mathcal{U}(\mathfrak{osp}(6|4))\rightarrow \mathcal{H}_{2|3}$ 
as follows:
\beq
\label{eq:representation of the algebra osp64}
\begin{split}
&\rho\left(
\begin{array}{cc|c|c|cc}
\textcolor{blue}{\mathfrak{L}^{\alpha}_{\beta}+\delta^{\alpha}_{\beta}\,\mathfrak{D}} &
\mathfrak{Q}^a_{\alpha} &
 -\bar{\mathfrak{q}}_{\alpha}  &  -\mathfrak{q}_{\alpha} & \textcolor{blue}{\mathfrak{P}_{\alpha\beta}} & \mathfrak{Q}^+_{\alpha b} \\
\mathfrak{S}_{a}^{\beta} & \textcolor{red}{-\mathfrak{R}^a_{b}-\delta^a_{b}\,\mathfrak{U}} &
 \textcolor{red}{\bar{\mathfrak{r}}_a} & \textcolor{red}{\mathfrak{r}_a} & \mathfrak{Q}^+_{\beta a} &\textcolor{red}{\epsilon_{ab}\,\mathfrak{r}}\\\hline
-\bar{\mathfrak{s}}^\beta & \textcolor{red}{\bar{\mathfrak{t}}^b} & \textcolor{red}{\mathfrak{u}} & 0 
&\mathfrak{q}_\beta& \textcolor{red}{-\mathfrak{r}_b} \\\hline
-\mathfrak{s}^\beta& \textcolor{red}{\mathfrak{t}^b}& 0& \textcolor{red}{-\mathfrak{u}} & 
\bar{\mathfrak{q}}_\beta& \textcolor{red}{-\bar{\mathfrak{r}}_b} \\\hline
\textcolor{blue}{\mathfrak{K}^{ \alpha \beta}} & \mathfrak{S}_-^{\alpha b} & -\mathfrak{s}^\alpha & -\bar{\mathfrak{s}}^\alpha & \textcolor{blue}{-(\mathfrak{L}^t)_{\alpha}^{\beta}-\delta_{\alpha}^{\beta}\,\mathfrak{D}} & 
(\mathfrak{S}^t)_{\alpha }^{b} \\
\mathfrak{S}_-^{\beta a} & \textcolor{red}{\epsilon^{ab}\,\mathfrak{t}} & \textcolor{red}{\mathfrak{t}^a} & \textcolor{red}{\bar{\mathfrak{t}}^a}& -(\mathfrak{Q}^t)_{a}^{\beta} & \textcolor{red}{(\mathfrak{R}^t)_{a}^{b}+\delta_{a}^{b}\,\mathfrak{U}}  \\
\end{array}
\right)=\,
\\
&\qquad \qquad \,=\,\left(
\begin{array}{cc|c|c|cc}
\textcolor{blue}{\frac{1}{2}\{\bar{\mathbf{a}}_{\beta},\mathbf{a}^{\alpha}\} }&
\bar{\mathbf{a}}_{\alpha}\mathbf{c}^a &
 -\bar{\mathbf{a}}_{\alpha}\bar{\mathbf{c}}  &  
-\bar{\mathbf{a}}_{\alpha}\mathbf{c} &
 \textcolor{blue}{\bar{\mathbf{a}}_{\alpha}\bar{\mathbf{a}}_{\beta}} &
\bar{\mathbf{a}}_{\alpha}\bar{\mathbf{c}}_{b}  \\
\bar{\mathbf{c}}_{a}\mathbf{a}^{\beta} & \textcolor{red}{-\frac{1}{2}[\bar{\mathbf{c}}_{b},\mathbf{c}^{a}] } &
 \textcolor{red}{\bar{\mathbf{c}}_{a}\bar{\mathbf{c}}} & \textcolor{red}{\bar{\mathbf{c}}_{a}\mathbf{c}} &
\bar{\mathbf{a}}_{\beta}\bar{\mathbf{c}}_{a} &
\textcolor{red}{\epsilon_{ab}\,\bar{\mathbf{c}} _1\bar{\mathbf{c}} _2}\\\hline
-\mathbf{a}^\beta  \mathbf{c}& \textcolor{red}{\mathbf{c}\mathbf{c}^b} &
 \textcolor{red}{[\mathbf{c},\bar{\mathbf{c}} ]} & 0 
&*& \textcolor{red}{*} \\\hline
-\bar{\mathbf{c}}\mathbf{a}^\beta& \textcolor{red}{\bar{\mathbf{c}}\mathbf{c}^b}& 0& \textcolor{red}{-[\mathbf{c},\bar{\mathbf{c}} ]} & 
*& \textcolor{red}{*} \\\hline
\textcolor{blue}{\mathbf{a}^{ \alpha}\mathbf{a}^{\beta}} &
\mathbf{a}^{\alpha}\mathbf{c}^{b} &* & * & 
\textcolor{blue}{*} & *\\
\mathbf{a}^{\beta}\mathbf{c}^{a} & \textcolor{red}{\epsilon^{ab}\,\mathbf{c}^1\mathbf{c}^2} &
 \textcolor{red}{*} & \textcolor{red}{*}& * & * \\
\end{array}
\right)\,.
\end{split}
\eeq
Notice that the entries $*$ are redundant, so we did not write them explicitly.
We have the following generators of the different $\g_{(a,b)}$ subalgebras:
\beq
\g_{(-,-)}=\text{span}\{\mathfrak{K}^{\alpha\beta}, \mathfrak{S}_-^{\alpha a }, \mathfrak{t}\}\,,\qquad 
\g_{(-,0)}=\text{span}\{\mathfrak{s}^{\alpha},\mathfrak{t}^a\}\,,\qquad
\g_{(0,-)}=\text{span}\{\bar{\mathfrak{s}}^{\alpha},\bar{\mathfrak{t}}^a\}\,,
\eeq
for the negative part $\g_{<0}$, 
\beq
\g_{(0,0)}=\text{span}\{\mathfrak{L}^{\alpha}_{\phantom{\alpha}\beta},\mathfrak{D},\mathfrak{R}^{a}_{\phantom{a}b},\mathfrak{U},\mathfrak{u}, \mathfrak{Q}^a_{\alpha}, \mathfrak{S}_{a}^{\alpha}\}\,,
\eeq
for $\g_{(0,0)}$ and finally for the positive $\g_{>0}$ one:
\beq
\g_{(+,0)}=\text{span}\{\mathfrak{q}_{\alpha},\mathfrak{r}_a\}\,,\qquad
\g_{(0,+)}=\text{span}\{\bar{\mathfrak{q}}_{\alpha},\bar{\mathfrak{r}}_a\}\,,\qquad
\g_{(+,+)}=\text{span}\{\mathfrak{P}_{\alpha\beta}, \mathfrak{Q}^+_{\alpha a}, \mathfrak{r}\}\,.
\eeq
It is convenient to introduce the combinations
\beq\label{JLRdef}
\mathfrak{C}\,=\,\mathfrak{D}+\mathfrak{U}\,,
\qquad
\mathfrak{C}_{L/R}\,=\,\mathfrak{D}+\mathfrak{J}_{L/R}\,,
\qquad
\mathfrak{J}_{L/R}\,=\,-\mathfrak{U}\pm\frac{1}{2}\,\mathfrak{u}\,,
\qquad
\mathfrak{B}\,=\,\mathfrak{D}-\mathfrak{U}\,.
\eeq
The generator $\mathfrak{C}$ gives the central charge of the $\mathfrak{su}(2|2)$ subalgebra. 
The $\mathfrak{so}(6)$ Dynkin labels $(r_1,r_2,r_3)$ are the eigenvalues of  the Cartan generators $(H_1,H_2,H_3)$ 
on the highest weight state.
Their expressions in our basis can be read off from 
\beq
\rho(H_i)\,=\,\frac{1}{2}[\mathbf{c}^i,\bar{\mathbf{c}}_i]\,
\qquad i=1,2,3\,.
\eeq
In particular $\mathfrak{U}=-\tfrac{1}{2}(H_1+H_2)$.
We conclude that, if $\hws$ is the highest weight vector of a representation of $\mathfrak{osp}(6|4)$, then the action of the Cartan subalgebra on $\hws$ is given by
\begin{align}
\label{hwsCandu}
&\mathfrak{C}\,\hws\,=\,
(\Delta-\tfrac{1}{2}(r_1+r_2))\,\hws\,=\,
c_{2|2}\,\hws\,, &
& \mathfrak{u}\,\hws\,=\,
2r_3\,\hws\,, & & &\nonumber\\
&\mathfrak{C}_{L/R}\,\hws\,=\,2b_{\pm }\,\hws\,,& &\mathfrak{L}_{1}^1\,\hws\,=\,s\,\hws\,,& &\mathfrak{R}_{1}^1\,\hws\,=\,j\,\hws\,.&
\end{align}

\subsection{Triangular decomposition of $OSP(6|4)$}
\label{app: triangular decomposition of osp}

Let us now describe the subgroups corresponding to the decomposition \eqref{ZZgraded} of the group $OSP(6|4)$ using the explicit realization \eqref{eq:representation of the algebra osp64}.

\beq\label{Groups_triang:APP}
\begin{split}
G_{(-,-)}&=\Bigg{\{}
g\,=\,\text{\footnotesize{$
\left(\begin{array}{c|cc|c}
\mathbbm{1}_4 &  &  & \\\hline
& 1 &  & \\
 &  & 1 & \\\hline
C &  &  & \mathbbm{1}_4 
\end{array}\right)$}}
\,\Big{|}\,\,
C^{st}\,\Sigma+C\,=\,0
\Bigg{\}}\,,\\
G_{(-,0)}&=\Bigg{\{}
g\,=\,\text{\footnotesize{$
\left(\begin{array}{c|cc|c}
\mathbbm{1}_4 &  &  & \\\hline
& 1 &  & \\
c &  & 1 & \\\hline
& c^{st}  &  & \mathbbm{1}_4 
\end{array}\right)$}}\,\,
\Bigg{\}}\,,
\qquad 
G_{(0,-)}=\Bigg{\{}
g\,=\,\text{\footnotesize{$
\left(\begin{array}{c|cc|c}
\mathbbm{1}_4 &  &  & \\\hline
\bar{c}& 1 &  & \\
 &  & 1 & \\\hline
&  &  \bar{c}^{st} & \mathbbm{1}_4 
\end{array}\right)$}}\,\,
\Bigg{\}}\,,
\\
G_{(0,0)}&=\Bigg{\{}
g\,=\,\text{\footnotesize{$
\left(\begin{array}{c|cc|c}
A &  &  & \\\hline
 & a_{11} & a_{12}  & \\
 & a_{21} & a_{22} & \\\hline
& & &(A^{st})^{-1} 
\end{array}\right)$}}\,\,
\,\Big{|}\,\,
A\,\in\,GL(2|2)\,,
\,\,\left(\begin{smallmatrix}a_{11}&a_{12}\\a_{21}&a_{22}\end{smallmatrix}\right)\,\in\,O(2)
\Bigg{\}}\,,\\
G_{(0,+)}&=\Bigg{\{}
g\,=\,\text{\footnotesize{$
\left(\begin{array}{c|cc|c}
\mathbbm{1}_4 & \bar{b}  &  & \\\hline
& 1 &  & \\
 &  & 1 &  \bar{b}^{st} \\\hline
&  &  & \mathbbm{1}_4 
\end{array}\right)$}}\,\,
\Bigg{\}}\,,
\qquad 
G_{(+,0)}=\Bigg{\{}
g\,=\,\text{\footnotesize{$
\left(\begin{array}{c|cc|c}
\mathbbm{1}_4 &  &  b & \\\hline
& 1 &  & b^{st} \\
 &  & 1 & \\\hline
&  && \mathbbm{1}_4 
\end{array}\right)$}}\,\,
\Bigg{\}}\,,\\
G_{(+,+)}&=\Bigg{\{}
g\,=\,\text{\footnotesize{$
\left(\begin{array}{c|cc|c}
\mathbbm{1}_4 &  &  & B\\\hline
& 1 &  &  \\
 &  & 1 & \\\hline
 &  &  & \mathbbm{1}_4 
\end{array}\right)$}}
\,\Big{|}\,\,
\Sigma\,B^{st}+B\,=\,0
\Bigg{\}}\,,
\end{split}
\eeq
The empty entries of theses matrices are zero.
Explicitly, the condition $\left(\begin{smallmatrix}a_{11}&a_{12}\\a_{21}&a_{22}\end{smallmatrix}\right)\,\in\,O(2)$ 
reads 
$a^t\sigma_1\,a\,=\,\sigma_1$.
We denote by $\widetilde{G}_{(0,0)}$ the subgroup of $G_{(0,0)}$  of elements with  superdeterminant one, 
more explicitly
 \beq\label{Gtilde00}
 \widetilde{G}_{(0,0)}=\Bigg{\{}
g\,=\,\text{\footnotesize{$
\left(\begin{array}{c|cc|c}
A &  &  & \\\hline
& a^{+1}&  & \\
 & & a^{-1} & \\\hline
& & &(A^{st})^{-1} 
\end{array}\right)$}}\,\,
\,\Big{|}\,\,
A\,\in\,GL(2|2)\,,
\,\,a\,\in\,GL(1)
\Bigg{\}}\,,
 \eeq
Notice that supertransposition maps $G_{(a,b)}$ to $G_{(-a,-b)}$.
The automorphism $g\mapsto \eta^{-1}g\eta$  maps $G_{(a,b)}$ to $G_{(-b,-a)}$.
Let us now give the explicit form of the transformations that we omitted in Table \ref{tab:Xpmtransf}.
We have the transformation properties
\beq
\begin{split}
\label{CtransfApp1}
g\circ(\mathcal{X}_+,W)\,=\,&
((1+\mathcal{X}_+C)^{-1}\mathcal{X}_+,(1+\mathcal{X}_+C)^{-1}W)\,,
\\
g\circ(\mathcal{X}_-,\bar{W})\,=\,&
((1+\mathcal{X}_-C)^{-1}\mathcal{X}_-,(1+\mathcal{X}_-C)^{-1}\bar{W})\,,
\end{split}
\eeq
for $g\,\in\,G_{(-,-)}$.
The transformations  $g\,\in\,G_{(+,+)}$ are 
\beq
\label{CtransfApp2}
g\circ(\mathcal{X}_+,W)\,=\,
(\mathcal{X}_++B,W)\,,\qquad 
g\circ(\mathcal{X}_-,\bar{W})\,=\,
(\mathcal{X}_-+B,\bar{W})\,.
\eeq

\section{Transformation properties and projectors}
\label{App:proofs}

In this appendix we wish to present the derivation of the transformation properties of certain important objects appearing in the main text. We also present the explicit expressions for the projectors appearing in the correlator of highest weight state of the stress-tensor multiplet.

\subsection{Covariant combinations from the coset construction}
\label{App:covariants_fromcoset}

It follows from 
\eqref{coset_action} that
the combination 
\beq
D_{12}\equiv D_{12}(\mathbf{p}_1,\mathbf{p}_2)\colonequals E(\mathbf{p}_2)^{-1}E(\mathbf{p}_1)\,=\,
\text{\footnotesize{$\left(\begin{array}{c|cc|c}
\mathbbm{1}_4  & \bar{W}_{\bar 1\bar2} & W_{12} & \mathbf{X}_{\mathbf{1}\tilde{\mathbf{2}}} \\\hline
0 & 1 & 0 & W_{12}^{st} \\
0 & 0 & 1 & \bar{W}_{\bar1\bar2}^{st}\\\hline
0 & 0 & 0 &  \mathbbm{1}_4  
\end{array}\right)$}}\,,
\eeq
transforms as
\beq\label{Dtransf}
g\circ D_{12}\,=\,E(g\circ\mathbf{p}_2)^{-1}E(g\circ\mathbf{p}_1)\,=\,
\left(g^*_{\mathbf{p}_2}\right)^{+1}
\,D_{12}\,
\left(g^*_{\mathbf{p}_1}\right)^{-1}\,.
\eeq
Let us notice that the action \eqref{coset_action}  can be extended to $g\,\in\,GL(6|4)$. 
It follows that $\mathbf{X}_{\mathbf{1}\tilde{\mathbf{2}}}$ is covariant under $GL(6|4)\supset OSP(6|4)$.
The matrix  $g^*_{\mathbf{p}}$, defined in \eqref{coset_action}, takes the form 
\beq
\label{Hdef:app}
g^*_{\mathbf{p}}\,=\,
\text{\footnotesize{$
\left(\begin{array}{c|cc|c}
H_{\mathbf{p}}^{}  & 0 &0  & 0 \\\hline
\ell_{\mathbf{p}}^{}  & k_{\mathbf{p}}^{+1}&  0& 0 \\
\bar{\ell}_{\mathbf{p}}^{}   &  0 & k_{\mathbf{p}}^{-1} & 0 \\\hline
* & \bar{m}_{\mathbf{p}}^{}  & m_{\mathbf{p}}^{}   &(H_{\mathbf{p}}^{st})^{-1}
\end{array}\right)$}}\,.
\eeq
Since $g^*_{\mathbf{p}}\in OSP(6|4)$ we have the conditions
$\bar{\ell}_{\mathbf{p}}^{st}\,k_{\mathbf{p}}=H_{\mathbf{p}}^{st}\,\bar{m}_{\mathbf{p}}$, 
$\ell_{\mathbf{p}}^{st}\,k_{\mathbf{p}}^{-1}=H_{\mathbf{p}}^{st}\,m_{\mathbf{p}}$.
The transformation \eqref{Dtransf} implies that
\beq
\begin{split}
W_{12}&\mapsto H_{\mathbf{p}_2} W_{12}k_{\mathbf{p}_1}-H_{\mathbf{p}_2}\mathbf{X}_{\mathbf{1}\tilde{\mathbf{2}}} \ell_{\mathbf{p}_1}^{st}=H_{\mathbf{p}_1} W_{12}k_{\mathbf{p}_2}+H_{\mathbf{p}_1}\mathbf{X}_{\mathbf{2}\tilde{\mathbf{1}}} \ell_{\mathbf{p}_2}^{st}\,,\\
\bar{W}_{\bar{1}\bar{2}}&\mapsto H_{\mathbf{p}_2} \bar{W}_{\bar{1}\bar{2}}k_{\mathbf{p}_1}^{-1}-H_{\mathbf{p}_2}\mathbf{X}_{\mathbf{1}\tilde{\mathbf{2}}} \bar{\ell}_{\mathbf{p}_1}^{st}=H_{\mathbf{p}_1} \bar{W}_{\bar{1}\bar{2}}k_{\mathbf{p}_2}^{-1}+H_{\mathbf{p}_1}\mathbf{X}_{\mathbf{2}\tilde{\mathbf{1}}} \bar{\ell}_{\mathbf{p}_2}^{st}\,,
\end{split}
\eeq
together with \eqref{X1tilde2_transf}, i.e. $\mathbf{X}_{\mathbf{1}\tilde{\mathbf{2}}}\mapsto H_{\mathbf{p}_2}\mathbf{X}_{\mathbf{1}\tilde{\mathbf{2}}}H_{\mathbf{p}_1}^{st}$.

We have for the non-trivial transformations in $G_{<0}$ the following expressions for $g_{\mathbf{p}}^*$. First for $g\,\in\,G_{(-,0)}$, 
\beq
\label{eq:gstarforG-0}
H_{\mathbf{p}}=h_{p}\,,\quad  
 k_{\mathbf{p}}=(1+cW_{{p}})^{-1}\,,  \quad \ell_{\mathbf{p}}=0\,, \quad  \bar{\ell}_{\mathbf{p}}=(1+cW_{{p}})cH_{\mathbf{p}}\,,
\eeq
then for $g\,\in\,G_{(0,-)}$
\beq
\label{eq:gstarforG0-}
H_{\mathbf{p}}=\bar{h}_{\bar{p}}\,,\quad 
 k_{\mathbf{p}}=(1+\bar{c}\bar{W}_{\bar{p}})\,,\quad  \ell_{\mathbf{p}}=(1+\bar{c}\bar{W}_{\bar{p}})\bar{c}H_{\mathbf{p}}\,,\quad   \bar{\ell}_{\mathbf{p}}=0\,,
\eeq
and finally for $g\,\in\,G_{(-,-)}$
\begin{align}
&H_{\mathbf{p}}=\left(1-\Sigma X_+^{st}C \right)^{-1}\,,& &k_{\mathbf{p}}=1-W_{p}(C^{-1}+\mathcal{X}_{+,\mathbf{p}})^{-1}\bar{W}_{\bar{p}}\,,& \nonumber \\ &\ell_{\mathbf{p}}=-k_{\mathbf{p}}W_{p}^{st}CH_{\mathbf{p}}\,,&  &\bar{\ell}_{\mathbf{p}}=-k_{\mathbf{p}}^{-1}\bar{W}_{\bar{p}}^{st}CH_{\mathbf{p}}\,,&
\end{align}
where $h_{\mathbf{p}}$, $\bar{h}_{\mathbf{p}}$ were defined around \eqref{eq:transformationpropertiesofthemathcalX}, $X_+$ around \eqref{eq:definitionofcosetrepresentative} and $m_{\mathbf{p}}$, $\bar{m}_{\mathbf{p}}$ follow from the above due to the constraint $g^*_{\mathbf{p}}\in OSP(6|4)$. We refer to \eqref{Groups_triang:APP} for the way $c$, $\bar{c}$ and $C$ enter the parametrization of $G_{<0}$.

\subsection{Covariance properties of  the three-point object \texorpdfstring{$\Lambda_{1,\bar{2},\mathbf{3}}$}{Lambda}}
\label{App:covariance_Lambda}

In the following, we want to derive the covariant $\Lambda_{1,\bar{2},\mathbf{3}}$ given in \eqref{eq:definitionLambda123}
 by applying superconformal transformations in order to go to special frames.
 Let us take three points of the kind $\{1, \bar{2}, \mathbf{3}\}$ and act on them with the superconformal group. Just by using $G_{>0}$, we can send them to $\{(\mathcal{X}_{1\bar 3},W_{13}), (-\mathcal{X}_{3\bar 2},\bar{W}_{\bar2\bar3}), 0\}$. We can now follow two strategies. Either, we first act with $G_{(-,0)}$ to send $\bar{W}_{\bar2\bar3}$ to zero and then with $G_{(0,-)}$ to send $W_{13}$ to zero, or we perform the operations in the opposite order. In the first case, we get $\{(H_{1\bar2\mathbf{3}}\mathcal{X}_{1\bar 3}H_{1\bar2\mathbf{3}}^{st},0), (-\mathcal{X}_{3\bar 2},0), 0\}$, while in the second $\{(\mathcal{X}_{1\bar 3},0), (-\bar{H}_{1\bar2\mathbf{3}}\mathcal{X}_{3\bar 2}\bar{H}_{1\bar2\mathbf{3}}^{st},0), 0\}$, where we have defined the matrices
\beq
\label{eq:def3pointHandbarH}
H_{1\bar2\mathbf{3}}\colonequals \big(1+W_{13}\bar{W}_{\bar2\bar3}^{st}\mathcal{X}_{3\bar{2}}^{-1}\big)^{-1}\,,\qquad  \bar{H}_{1\bar2\mathbf{3}}\colonequals \big(1-\bar{W}_{\bar2\bar3}W_{13}^{st}\mathcal{X}_{1\bar{3}}^{-1}\big)^{-1}\,,
\eeq
obeying the intertwining identities
\beq
\label{eq:bridgeHidentities}
H_{1\bar2\mathbf{3}}(\mathcal{X}_{1\bar 2}-\mathcal{X}_{1\bar 3})H_{1\bar2\mathbf{3}}^{st}=\mathcal{X}_{3\bar 2}\,,\qquad \bar{H}_{1\bar2\mathbf{3}}(\mathcal{X}_{1\bar 2}-\mathcal{X}_{3\bar 2})\bar{H}_{1\bar2\mathbf{3}}^{st}=\mathcal{X}_{1\bar 3}\,.
\eeq
The formulas \eqref{eq:bridgeHidentities} essentially follow from the shift identity
\beq
\label{eq:mathcalX12asexpressedbyX13andX32}
\mathcal{X}_{1\bar{2}}=\mathcal{X}_{1\bar{3}}+\mathcal{X}_{3\bar{2}}+W_{13}\wedge \bar{W}_{\bar 2\bar 3}\,.
\eeq
The identities \eqref{eq:bridgeHidentities} imply that the new matrices transform as $H_{1\bar2\mathbf{3}}\mapsto h_3 H_{1\bar2\mathbf{3}}h_1^{-1}$ under $G_{(-,0)}$ and $\bar{H}_{1\bar2\mathbf{3}}\mapsto \bar{h}_{\bar3} \bar{H}_{1\bar2\mathbf{3}}\bar{h}_{\bar2}^{-1}$ under $G_{(0,-)}$. This has also been checked directly.

Finally, for both of our strategies, we can send one of the remaining points to infinity using the $G_{(-,-)}$ action \eqref{CtransfApp1}. An appropriate choice of the point to send to infinity, sends in both cases the remaining point to $\Lambda^{-1}_{1\bar 2,\mathbf{3}}$, where we have defined
\beq
\label{eq:definitionLambda123APP}
\begin{split}
\Lambda_{1\bar 2,\mathbf{3}}&\colonequals
\big(H_{1\bar2\mathbf{3}}\mathcal{X}_{1\bar 3}H_{1\bar2\mathbf{3}}^{st}\big)^{-1}+\mathcal{X}_{3\bar 2}^{-1}=
\mathcal{X}_{1\bar 3}^{-1}+\big(\bar{H}_{1\bar2\mathbf{3}}\mathcal{X}_{3\bar 2}\bar{H}_{1\bar2\mathbf{3}}^{st}\big)^{-1}\\
&=\mathcal{X}_{1\bar 3}^{-1}+\mathcal{X}_{3\bar 2}^{-1}-\left(\mathcal{X}_{1\bar 3}^{-1}W_{13}\right)\wedge \left(\mathcal{X}_{3\bar 2}^{-1}\bar{W}_{\bar 2\bar 3}\right)\Sigma
=
(\bar{H}_{1\bar2\mathbf{3}}^{st})^{-1}
 \mathcal{X}_{3\bar 2}^{-1}
 \mathcal{X}_{1\bar 2}^{}
 \bar{H}_{1\bar2\mathbf{3}}^{st}
\mathcal{X}_{1\bar 3}^{-1}\,.
\end{split}
\eeq
where in the last equality we have used \eqref{eq:bridgeHidentities}.
From the definition of $\Lambda_{1\bar 2,\mathbf{3}}$ and the transformation properties of $H_{1\bar 2\mathbf{3}}$ and $\bar{H}_{1\bar 2\mathbf{3}}$, the transformation properties \eqref{Lambdatransf} follow. To then obtain \eqref{eq:definitionLambda123} in the main text, simply use the fact that $ \mathcal{G}_{1\bar{2},\mathbf{3}}= (\bar{H}_{1\bar2\mathbf{3}}^{st})^{-1}$ in \eqref{eq:definitionLambda123APP}.

\subsection{Chiral structures}
\label{App:chiral structures}

 It follows from \eqref{eq:actionofomega} and \eqref{Otransf} that chiral correlators are invariant under  
  $G_{\text{easy}}\times G_{(0,+)}$ where
 $G_{\text{easy}}:=G_{(+,+)}\times G_{(+,0)}\times G_{(0,-)}$. 
 It is not hard to verify that invariance under $G_{\text{easy}}$  implies that  correlation functions of chiral operators 
 can depend only on the combinations
  \beq\label{Invforchiral}
 \mathcal{X}_{ij,+}:=\mathcal{X}_{i,+}-\mathcal{X}_{j,+}\,,
 \qquad
T_{i,jk}\,:=
\mathcal{X}_{ij,+}^{-1}W^{}_{ij}-\mathcal{X}_{ik,+}^{-1}W^{}_{ik}\,.
 \eeq
Analogous structures appear in the four dimensional case as well, see e.g.~\cite{Pickering:1999rk}, \cite{Dolan:2000uw}.
Given three points there is only one independent $T$ we can construct. This can be verified using the relations 
$\mathcal{X}_{13,+}\,T_{1,23}\,=\,\mathcal{X}_{23,+}\,T_{2,13}$, 
$T_{1,23}=-T_{1,32}$ and $T_{2,13}=T_{1,23}+T_{3,12}$
 which follow from the definition.
The constraint of  $G_{(0,+)}$ invariance is harder to solve. 
The latter,  combined with $G_{(0,-)}$ invariance, implies invariance under a subgroup of  $\widetilde{G}_{(0,0)}$
corresponding to  $a^{-1}=\text{sdet} \,A$.
Invariance under this subgroup is strong enough to imply that the three point function of chiral operators
 is independent of $T_{1,23}$.
Before proving this statement let us notice that it immediately implies that the three-point function of chiral operators vanishes,
as $T_{1,23}$ is the only combination in  \eqref{Invforchiral} that carries non trivial $GL(1)_a\subset \widetilde{G}_{(0,0)}$, see \eqref{Gtilde00},  weight and the three-point function carries positive weight under this action.

We will now show that any invariant under the subgroup of $\widetilde{G}_{(0,0)}$ corresponding to $a^{-1}=\text{sdet} \,A$
constructed out of $\mathcal{X}_{12,+},\mathcal{X}_{13,+}$ and $T:=T_{1,23}$
is independent of $T$. To do so it is convenient to go to a frame where 
$\mathcal{X}_{12,+}=\Omega ,\mathcal{X}_{13,+}=\Omega\,D$ and
 $T=\left(\begin{smallmatrix}\bar{\xi}^{\alpha}\\\bar{v}^a\end{smallmatrix}\right)$,
where $\Omega$ is given in \eqref{Omegadef} and 
$D:=\text{diag}(z_+,\bar{z}_+,\omega_+,\omega_+)$.
It is then easy to see that invariance under the  group   $SP(2)\times GL(1)\times D_4$ defined below \eqref{framemapTEXTtwofull2} that acts on this class of frames implies that the required invariant is independent of $T$. 
\subsection{Covariance properties of \texorpdfstring{$\mathcal{Z}_{1\bar 23\bar 4}$}{Z}}
\label{App:covariance_monodromyNEW}

In this part of the appendix, we wish to describe how to obtain the covariant matrix $ \calZ_{1\bar 23\bar 4}$ appearing in \eqref{Zfulldef} and how to show that it its eigenvalues are invariant.

 We take four generic points $\{1,\bar 2, 3,\bar 4\}$ and by applying a $G_{>0}$ transformation bring them in the form $\{0,(-\mathcal{X}_{1\bar 2},0),(-(\mathcal{X}_{1\bar 2}-\mathcal{X}_{3\bar 2}),W_{31}),(-\mathcal{X}_{1\bar 4},\bar{W}_{\bar 4\bar 2})\}$. We can then apply  a $G_{(0,-)}$ transformation to set $W_{31}$ to zero, a $G_{(-,0)}$ one to set $\bar{W}_{\bar 2}'=\bar{W}_{\bar 4}'$ and finally a $G_{(0,+)}$ one to set $\bar{W}_{\bar 2}'=\bar{W}_{\bar 4}'=0$. We are then left with the points $\{0,(-\mathcal{X}_{1\bar 2},0),(-(\mathcal{X}_{1\bar 2}-\mathcal{X}_{3\bar 2}),0),(-\bar{H}_{1\bar 2,3\bar 4}^{-1}\mathcal{X}_{1\bar 4}(\bar{H}_{1\bar 2,3\bar 4}^{-1})^{st},0)\}$,  where $\bar{H}_{1\bar 2,3\bar 4}\colonequals 1+\bar{W}_{\bar 2 \bar 4}W_{13}^{st}(\mathcal{X}_{1\bar 2}-\mathcal{X}_{3\bar 2})^{-1}$. The matrix $\bar{H}_{1\bar 2,3\bar 4}$ is a generalization of $\bar{H}_{1\bar 2\mathbf{3}}$ of \eqref{eq:def3pointHandbarH} and obeys the intertwining relation
\beq
\label{eq:4pointsbridge}
\bar{H}_{1\bar 2,3\bar 4}(\mathcal{X}_{1\bar 2}-\mathcal{X}_{3\bar 2})\bar{H}_{1\bar 2,3\bar 4}^{st}=\mathcal{X}_{1\bar 4}-\mathcal{X}_{3\bar 4}\,,
\eeq
 due to the four point identity $\mathcal{X}_{1\bar 2}-\mathcal{X}_{3\bar 2}-\mathcal{X}_{1\bar 4}+\mathcal{X}_{3\bar 4}=W_{13}\wedge \bar{W}_{\bar 2\bar 4}$. Hence, it transforms as $\bar{H}_{1\bar 2,3\bar 4}\mapsto \bar{h}_{\bar 4}\bar{H}_{1\bar 2,3\bar 4}\bar{h}_{\bar 2}^{-1}$ under $G_{(0,-)}$ transformations. 
  Finally, we can use  a $G_{(-,-)}$ transformation to send the fourth point to infinity, bringing the points to the form
\beq\label{framemap}
\Big{\{}1,\bar{2},3,\bar{4}
\Big{\}}\,
\mapsto\,
\Big{\{}0,(\mathcal{Z}_-,\,0),(\mathcal{Z}_+,\,0),(\infty,0)
\Big{\}}\,,
\eeq
where we have defined
\beq
\label{eq:definitionZpm}
\begin{split}
\mathcal{Z}_-&\colonequals -(1-\mathcal{X}_{1\bar 2}\bar{H}_{1\bar 2,3\bar 4}^{st}\mathcal{X}_{1\bar 4}^{-1}\bar{H}_{1\bar 2,3\bar 4})^{-1}\mathcal{X}_{1\bar 2}\,,\\
\mathcal{Z}_+&\colonequals-(1-(\mathcal{X}_{1\bar 2}-\mathcal{X}_{3\bar 2})\bar{H}_{1\bar 2,3\bar 4}^{st}\mathcal{X}_{1\bar 4}^{-1}\bar{H}_{1\bar 2,3\bar 4})^{-1}(\mathcal{X}_{1\bar 2}-\mathcal{X}_{3\bar 2})\,,
\end{split}
\eeq
Thanks to the transformation properties of $\bar{H}_{1\bar 2,3\bar 4}$, it is clear that the object
 \beq
 \label{eq:definitionofcalZ1234app}
 \calZ_{1\bar 23\bar 4}\colonequals \mathcal{X}_{1\bar 2}(1-\mathcal{Z}_-^{-1}\mathcal{Z}_+)^{-1}\mathcal{X}_{1\bar 2}^{-1}=\mathcal{X}_{1\bar 2}\calN\mathcal{X}_{1\bar 4}^{-1} \mathcal{X}_{3\bar 4}\calN^{-1}\mathcal{X}_{3\bar 2}^{-1}\,,
 \eeq
 transforms covariantly under the action of $G_{(0,-)}$. In \eqref{eq:definitionofcalZ1234app} we have used the definition \eqref{Zfulldef} from which follows $\calN=\bar{H}_{1\bar 2,3\bar 4}^{st}$. 
 
   To show that $ \calZ_{1\bar 23\bar 4}$ transforms covariantly also under $G_{(-,0)}$ demands a bit more work. We first want to prove that 
 \beq
 \label{eq 4Has function of 3H}
 \bar{H}_{1\bar 2,3\bar 4}=\bar{H}_{1\bar 4\mathbf{3}}^{-1}\bar{H}_{1\bar 2\mathbf{3}}\ \Longrightarrow\ \calN=\mathcal{G}_{1\bar{2},\mathbf{3}}\mathcal{G}_{1\bar{4},\mathbf{3}}^{-1}\,.
 \eeq
 To show the above, we first observe that due to \eqref{eq:bridgeHidentities}, $\bar{H}_{1\bar 2\mathbf{3}}$ can be written as $\bar{H}_{1\bar 2\mathbf{3}}=1+\bar{W}_{\bar 2\bar 3}W_{13}^{st}(\mathcal{X}_{1\bar 2}-\mathcal{X}_{3\bar 2})^{-1}$. Using the shift relation \eqref{eq:mathcalX12asexpressedbyX13andX32} and the graded antisymmetry of $\mathcal{X}_{1\bar 3}^{-1}$, we can expand $\bar{H}_{1\bar 2,3\bar 4}$ in a geometric power series
 \beq
 \label{eq:tempexpansionofbarH}
  \bar{H}_{1\bar 2,3\bar 4}=1+\bar{W}_{\bar 2\bar 4}W_{13}^{st}\mathcal{X}_{1\bar 3}^{-1}\sum_{n=0}^{\infty}\left(\bar{W}_{\bar 2\bar 3}W_{13}^{st}\mathcal{X}_{1\bar 3}^{-1}\right)^n\,.
 \eeq
 In a similar fashion, we find 
 \beq
\bar{H}_{1\bar 4\mathbf{3}}^{-1}\bar{H}_{1\bar 2\mathbf{3}}=\left(1-\bar{W}_{\bar 4\bar 3}W_{13}^{st}\mathcal{X}_{1\bar 3}^{-1}\right)\left(1+\bar{W}_{\bar 2\bar 3}W_{13}^{st}(\mathcal{X}_{1\bar 2}-\mathcal{X}_{3\bar 2})^{-1}\right)\,,
 \eeq 
 which is easily seen to be equal to \eqref{eq:tempexpansionofbarH} after using \eqref{eq:mathcalX12asexpressedbyX13andX32} to expand $(\mathcal{X}_{1\bar 2}-\mathcal{X}_{3\bar 2})^{-1}$ in a geometric power series in $\mathcal{X}_{1\bar 3}^{-1}$. We observe that equation \eqref{eq 4Has function of 3H} requires the use of the anti-chiral point $\bar{3}$, but its effects on the eigenvalues of $\calZ_{1\bar 23\bar 4}$ will cancel in the end. Note that the relation \eqref{eq 4Has function of 3H} is compatible with the intertwining relations \eqref{eq:4pointsbridge} and \eqref{eq:bridgeHidentities}. 
 It follows from \eqref{eq 4Has function of 3H} and the definition \eqref{eq:definitionLambda123} that
\beq
\Lambda_{1\bar 4\mathbf{3}}^{-1}\Lambda_{1\bar 2\mathbf{3}}=\left(\mathcal{X}_{1\bar3} \mathcal{G}_{1\bar{2},\mathbf{3}}\mathcal{X}_{1\bar2}^{-1}\right)\calZ_{1\bar 23\bar 4}\left(\mathcal{X}_{1\bar3} \mathcal{G}_{1\bar{2},\mathbf{3}}\mathcal{X}_{1\bar2}^{-1}\right)^{-1}\,.
\eeq
Hence $\calZ_{1\bar 23\bar 4}$ is similar to a covariantly transforming matrix and hence its eigenvalues are invariant.

We can always use the groups symmetry to set the Gra\ss mann variables to zero. In that case, it becomes easy to see that $\calZ_{1\bar 23\bar 4}$ takes the form 
\beq
\calZ_{1\bar 23\bar 4}\big{|}_{\text{ferm}=0}\,=\,\begin{pmatrix}
x_{12}x_{14}^{-1}x_{34}x_{32}^{-1} & 0 \\
0 & \frac{y_{1\bar{2}}y_{3\bar{4}}}{y_{1\bar{4}}y_{3\bar{2}}}\,\mathbbm{1}_2
\end{pmatrix}\,.
\eeq
It follows that two of the eigenvalues of $\calZ_{1\bar 23\bar 4}$ are degenerate. Naming the eigenvalues of the conformal piece $z$ and $\bar z$, and the degenerate eigenvalue $w$, we find the three promised invariants appearing in \eqref{4points1_2BPS}.

\subsection{R-symmetry invariants and projectors}
\label{app:R-invariants}

In this short section we present the explicit expressions for the projectors appearing in \eqref{eq:Gforzerofermions}. Let us define the combination
\beq
\mathbf{w}_{1\bar 2 3 \bar 4}\colonequals\frac{y_{1\bar 2}y_{3\bar 4}}{y_{1\bar 4}y_{3\bar 2}}\,.
\eeq
We can construct, up to inversions, $\binom{4}{2}=6$ such cross-ratios, namely
\begin{align}
&s_1\colonequals \mathbf{w}_{1\bar 2 4 \bar 3}\,,& &s_2\colonequals \mathbf{w}_{1\bar 3 2 \bar 4}\,,&
&s_3\colonequals \mathbf{w}_{1\bar 4 3 \bar 2}\,,&\nonumber\\
&\bar{s}_1\colonequals \mathbf{w}_{2\bar 1 3 \bar 4}\,,& &\bar{s}_2\colonequals \mathbf{w}_{3\bar 1 4 \bar 2}\,,&
&\bar{s}_3\colonequals \mathbf{w}_{4\bar 1 2 \bar 3}\,,&
\end{align}
subject to one algebraic relation,
$\prod_{j=1}^3s_j\bar{s}_j=1$\,. The irreducible representations appearing in the tensor product of two adjoints (remark that $\mathbf{15}\equiv [1,0,1]$ using SU(4) Dynkin labels) are
\be 
\mathbf{15} \times \mathbf{15} = \mathbf{1} + \mathbf{15} + \mathbf{20} + \mathbf{84} + \mathbf{15} + ( \mathbf{45}+ \mathbf{\bar{45}})\, ,
\ee
and the associated projectors take the following form:
\begin{align}
P_{\mathbf{1}} & = \frac{1}{15}\, ,
\nonumber\\\nonumber
P_{\mathbf{15}} & = \frac{1}{6}\left(-1 +\frac{1}{s_1} + \frac{1}{\bar{s}_1} +s_3 +  \bar{s}_3\right) \, , 
\\\nonumber
P_{\mathbf{20}} & = \frac{1}{12} -\frac{1}{8} \left(\frac{1}{s_1} + \frac{1}{\bar{s}_1} + s_3 + \bar{s}_3\right) +
 \frac{1}{4}\left( \frac{1}{s_1 \bar{s}_1} + s_3 \bar{s}_3 \right) 
 -\frac{1}{4} \left(\frac{1}{s_1 s_2 \bar{s}_1} + \frac{1}{\bar{s}_1 \bar{s}_2 s_1} \right)\, ,
\\\nonumber
P_{\mathbf{84}} & = \frac{1}{60} -\frac{1}{24} \left(\frac{1}{s_1} + \frac{1}{\bar{s}_1} + s_3 +\bar{s}_3 \right)
+ \frac{1}{4}\left(\frac{1}{s_1 \bar{s}_1} + s_3 \bar{s}_3 \right)
 + \frac{1}{4}\left( \frac{1}{s_1 \bar{s}_1 \bar{s}_2} + \frac{1}{\bar{s}_1 s_1 s_2}\right)\, ,
\\\nonumber
P_{\mathbf{15}} & = \frac{1}{8}\left(-\frac{1}{s_1} - \frac{1}{\bar{s}_1} + s_3 +\bar{s}_3 \right)
\\
P_{\mathbf{45}+ \mathbf{\bar{45}}} & =  \frac{1}{8}\left(\frac{1}{s_1} + \frac{1}{\bar{s}_1} - s_3 -\bar{s}_3 \right)
-\frac{1}{2}\left(\frac{1}{s_1 \bar{s}_1 } - s_3 \bar{s}_3\right)\, .
\end{align}

\newpage

\bibliographystyle{abe}
\bibliography{bibliography}

\end{document}